\documentclass[11pt,a4paper]{article}
\usepackage{amsmath,amssymb,bm,amsthm}
\usepackage[vcentermath,enableskew]{youngtab}
\usepackage{cite}
\usepackage{setspace}
\usepackage{fullpage}
\usepackage{graphicx}
\usepackage{amsfonts} 
\usepackage[colorlinks=true,linkcolor=red,citecolor=blue]{hyperref}   

\newcommand{\be}{\begin{equation}}
\newcommand{\ee}{\end{equation}}
\newcommand{\bear}{\begin{eqnarray}}
\newcommand{\ear}{\end{eqnarray}}
\newcommand{\nn}{\nonumber}

\newcommand{\al}{\alpha}
\newcommand{\bt}{\beta}
\newcommand{\gm}{\gamma}
\newcommand{\dl}{\delta}
\newcommand{\ep}{\epsilon}

\newcommand{\lm}{\lambda}
\newcommand{\ald}{\dot{\alpha}}
\newcommand{\btd}{\dot{\beta}}

\newcommand{\Au}{\underline{A}}
\newcommand{\Bu}{\underline{B}}
\newcommand{\Cu}{\underline{C}}
\newcommand{\n}{\raise4pt \hbox{\tiny{${\bf n}$}}}
\newcommand{\nmo}{ \raise4pt \hbox{\tiny{$\bf~\! n-1 $ }} }
\newcommand{\one}{\raise4pt \hbox{ \hspace{-4mm} \tiny{ $\bf 1$} }}
\newcommand{\two}{\raise4pt \hbox{\tiny{$\bf 2$}}}
\newcommand{\Cdots}{\raise1pt \hbox{\small{$\cdots$}}}
\newcommand{\CP}{\mathbb{C}P}
\newcommand{\C}{\mathbb{C}}

\newcommand{\g}{\mathfrak{g}}
\newcommand{\half}{\frac{1}{2}}
\renewcommand{\L}{\mathcal{L}}
\newcommand{\K}{\mathcal{K}}
\newcommand{\zb}{\bar{z}}
\newcommand{\del}{\partial}

\newcommand{\D}{D \!\!\!\!/\!~}
\renewcommand{\o}{| \Omega \rangle} 
\newcommand{\om}{| \omega \rangle}
\newcommand{\omb}{| \bar{\omega}\rangle}
\newcommand{\tr}{\mbox{tr} }
\newcommand{\Dots}{\cdot\cdot}
\newcommand{\vDots}{{\rotatebox{90}{$\Dots$}}}
\newcommand{\first}{1}
\newcommand{\kth}{k}
\newcommand{\I}{\mbox{\i}}
\newcommand{\J}{\mbox{\j}}

\title{A projective Dirac operator on $\CP^n$ and extended SUSY}
\author{ Idrish Huet\footnote{idrish@ifm.umich.mx}, Julieta Medina\footnote{jmedinag@ipn.mx}{ }  \\
\\
{$*$ \it Facultad de Ciencias en F\'{\i}sica y Matem\'aticas, Universidad Aut\'onoma de Chiapas,}\\
{\it Ciudad Universitaria, Tuxtla Guti\'errez 29050, M\'exico }\\
{} \\
{$\dagger$ \it Ciencias B\'asicas UPIITA-IPN, Av. IPN 2580,}\\
{\it Col. La Laguna Ticom\'an 07340, M\'exico CDMX Mexico}
\\
\\
\it Dedicated to the memory of Yula Cort\'es Bernatowich, friend and teacher}

\begin{document}
\maketitle
\begin{abstract}
We construct a universal spin$_c$ Dirac operator on $\CP^n$ built by projecting $su(n+1)$ left actions and prove its equivalence to the standard right action Dirac operator on $\CP^n$. The eigenvalue problem is solved and the spinor space constructed thereof, showing that the proposed Dirac operator is universal, changing only its domain for different spin$_c$ structures. Explicit expressions for the chirality and the eigenspinors are also found and consistency with the index theorem is established. Also the extended $\mathcal{N} =2$ supersymmetry algebra is realised through the Dirac operator and its companion supercharge, an expression for the superpotential of any spin$_c$ connection on $\CP^n$ is found and generalised to any any spin coset manifold $G/H$ with $G,H$ compact, connected, and $G$ semisimple. The $R$-symmetry of this superalgebra is found to be equivalent to the $U(1)$ holonomy of the spin$_c$ connection.
\end{abstract}
\section{Introduction}
\label{intro}
The complex projective planes, $\CP^n$, are noteworthy in several contexts within quantum field theory, for instance the 2-dimensional euclidean sigma models with $\CP^n$-valued fields and their solitonic solutions have been studied lately by Grundland and Post in \cite{Grundland}, and by Aguado, Asorey and Wipf in connection with the Nahm transform on the torus \cite{AAWipf}; it has also been pointed out that the $\mathcal{N}=2$ supersymmetric extensions of these sigma models provide non-trivial integrable models.  Developments in the last decade have shown that $\CP^n$ and its Dirac operator are also relevant to some models of the quantum Hall effect, where the square of the Dirac operator serves as a non-relativistic hamiltonian for the theory. In such models the spin degrees of freedom can be conceived as $\CP^n$-vaued fields, particularly for $n=1,3$; remarkably their excitations are topologically non-trivial solitons of $\CP^n$ \cite{Rajamaran}.  Having these applications in mind it seems important to have an understanding of the Dirac operator on $\CP^n$ and its many properties; progress has been done in this direction in \cite{Seifarth,Baer,Wipf,Universal,Huet,Smilga,Habib}. It is well known that $\CP^n$ is a symmetric riemannian space $G/H$, and hence there is a standard way to construct, at least in principle, the Dirac operator using the right actions of the Lie algebra of $G$ \cite{Bal,Fried}. This approach was also followed in \cite{Habib} where a bound was found for the spectrum of the twisted Dirac operator in K\"ahlerian submanifolds of $\CP^n$ for any spin$_c$ structure, more recent estimates were found for the eigenvalues of the Dirac operator on K\"ahler-Einstein manifolds in \cite{Nakad}.
 
It is a well known fact that right and left cosets $G/H, G \setminus \!H$ are equivalent, however left and right $G$ actions on a given, say left, coset are very different; in this paper we focus on the eigenvalue and spin$_c$ connection problem of the Dirac operator on $\CP^n$ seen as a coset space, from a non-standard point of view as in \cite{Huet}; namely finding a left-action Dirac operator. We formulate also the supersymmetry algebra of the extended supersymmetry that gives rise the Hamiltonian $H = \D^2$ of a non-relativistic fermion. It was shown in \cite{Huet} that the ansatz proposed there corresponds to the canonical spin$_c$ structure on $\CP^2$ without further clarification, in the present work we find the underlying reason and generalise the result in two ways: we deal with all spin$_c$ structures at once, and we treat any dimension $n$. Yet another rather interesting construction involving complex projective planes and their Dirac operator can be found in  \cite{Brian,BDolan} where it was shown that one fermion generation of the standard model can be obtained as a zero mode of the Dirac operator on $\CP^2 \times \CP^3$ and the possibility of obtaining three generations non-trivially was discussed. It is also undeniable that supersymmetry has acquired an important status amongst quantum field theories, the role of the extended supersymmetry on $\CP^n$ was treated elegantly in \cite{Wipf, susycpn}. In the spirit of the work done by Kirchberg, L\"ange and Wipf in \cite{Wipf} we present a realisation of the $\mathcal{N}=2$ supersymmetry algebra and superpotential in our approach, which is however quite different in that it makes explicit use of the coset structure and does not resort to local coordinates.

The paper is organised as follows: section 2 contains the definitions and conventions, section 3 presents the left action Dirac operator on $\CP^n$, and proves its equivalence to the standard right action operator. Section 4 treats the eigenvalue problem, we show the construction if the spinor space for any spin$_c$ structure. Section 5 deals with different realisations of the left-action chirality operator. Section 6 presents a group-theoretical treatment to find the structure of the vacuum state and the zero modes of the Dirac operator, which determine all eigenspinors. Section 7 shows how the $\mathcal{N} =2$ supersymmetry algebra is realised by the Dirac operator and its companion supercharge and presents our results regarding the superpotential and $R$-symmetry. Section 8 discusses the scope of the approach presented and extends some results to any spin coset space $G/H$ with $G,H$ compact and connected, and $G$ semisimple. 

\section{Conventions and definitions}
\label{sec:1}
In the following we want to use extensively the construction of the complex projective planes as the coset spaces $\CP^n = SU(n+1)/S(U(n) \times U(1))$. We will use a global coordinate system over $SU(n+1)$, proposed in \cite{CPNF}, to parametrise an element $g \in SU(n+1)$ in the fundamental representation by orthonormal complex vectors 
$u^{\al}_{\imath}$ and $z^{\al}$

\be \label{gcoord}
g = \left( \begin{array}{ccccc}
        u_1^1 & u^1_2 & \cdots & u^1_{n} & z^1 \\
        u_1^2 & u^2_2 & \cdots & u^2_{n} & z^2 \\
        \vdots & \vdots & \vdots & \vdots & \vdots \\
        u_1^{n+1} & u^{n+1}_2 & \cdots & u^{n+1}_{n} & z^{n+1} 
       \end{array}\right)
\ee
subjected to the constraints: $\bar{u}^{\imath}_{\al} u^{\al}_{\jmath} = \dl^{\imath}_{\jmath}$, $\bar{z}_\al u^{\al}_{\imath} = 0$ and $\zb_{\al} z^{\al} = 1$. Unless otherwise explicitly stated, we shall use Einstein's convention and the following ranges for indices:

\bear
\al, \bt &=& 1, \cdots, n+1 \nn \\
\imath, \jmath &=& 1, \cdots,n  \nn \\
\ald, \btd &\mbox{label}& s(u(n)\times u(1)) \nn \\
i,j &\mbox{label}& su(n+1)/s(u(n)\times u(1)) \nn \\
a,b &=& 1, \cdots, n^2 + 2n  \nn \\
A,B &=& n^2, n^2 + 1, \cdots, n^2 + 2n - 1 \nn \\
\Au, \Bu &=& 1,\cdots, n^2-1; n^2 + 2n \nn \\
I,J &=& 1, \cdots, n^2 -1
\ear
In choosing these ranges for indices we are actually using a diagonal block embedding $S(U(n)\times U(1)) \hookrightarrow SU(n+1)$. The indices $\ald, \btd, i,j$ label the isotropy subalgebra and its complement at each point of the coset space, so they do not take fixed values, but rather depend of the point through a rotation matrix.

The mentioned constraints may be expressed by the unitarity and orthogonality of $n$ columns along with either of the following conditions

\be \label{constr}
z^\al = \frac{1}{n!} \ep^{\bt_1 \cdots \bt_n \al}\ep_{\imath_1 \cdots \imath_n} \bar{u}^{\imath_1}_{\bt_1} \cdots \bar{u}^{\imath_n}_{\bt_n}, \quad
\bar{u}^{\imath}_{\al} = \frac{1}{(n-1)!}\ep_{\al \bt_1 \cdots \bt_n}\ep^{\imath \imath_1 \cdots \imath_{n-1}}u^{\bt_1}_{\imath_1} \cdots u^{\bt_{n-1}}_{\imath_{n-1}}z^{\bt_n} ~.
\ee
$\CP^n$ can be conceived as the $SU(n+1)$ adjoint orbit of $\mathcal{P}^0 = diag(0, \cdots, 0, 1)$, a rank one projector , following the construction put forward in \cite{CPNF}:

\be  \label{calP}
\mathcal{P} = \frac{{\bf 1}}{n+1} + \frac{\sqrt{n}}{\sqrt{2(n+1)}} \xi_a \lm_a   = g \mathcal{P}^0 g^{\dagger}.
\ee
Here $\lm_a$ are the standard Gell-Mann matrices which satisfy the matrix algebra

\be
\lm_a \lm_b = \frac{2}{n+1} \dl_{ab} {\bf 1} + (d_{abc} + i f_{abc})\lm_c,
\ee
and the coefficients $\xi_a$ are a global coordinate system for $\CP^n$, describing its embedding into $\mathbb{R}^{n^2+2n}$, we call ``north pole" the point where $\mathcal{P}=\mathcal{P}^0$; at the north pole the indices $i,j$ coincide with $A,B$ and $\ald, \btd$ with $\underline{A}, \underline{B}$. The coordinates are subjected to the constraints

\be
\xi_a \xi_a = 1, \qquad d_{abc}\xi_a \xi_b =  \frac{(n-1)\sqrt{2}}{\sqrt{n(n+1)}} \xi_c,
\ee
which are a consequence of $\mathcal{P}^2 = \mathcal{P}$. The metric and complex structure of $\CP^n$ in this coordinate system are known to be \cite{CPNF}

\be
P_{ab} = \frac{2}{n+1} \delta_{ab} + \frac{\sqrt{2n}}{\sqrt{n+1}} d_{abc} \xi_c - \frac{2n}{n+1} \xi_a \xi_b, \quad J_{ab} = \frac{\sqrt{2n}}{\sqrt{n+1}} f_{abc} \xi_c.
\ee
The metric is a projector onto the tangent space, $P^2 = P$, the complex structure is tangent $JP = PJ =J$, and satisfies $J^2 = -P$. The K\"ahler structure and its conjugate are then projectors onto the (anti)holomorphic tangent subspaces

\be
K^{\pm} = \frac{P \pm i J}{2}, \qquad (K^{\pm})^2 = K^{\pm}.
\ee
Using (\ref{calP}) and (\ref{gcoord}) one may obtain the following expression for the coordinates

\be \label{xiz}
\xi_a = \sqrt{\frac{n+1}{2n}} \zb_\al (\lm_a)^\al_\bt z^\bt = -\half \tr ( g \lm_{n^2 +2n} g^{\dagger} \lm_a)~.
\ee
The isotropy subalgebra of $\mathcal{P}^0$ is generated by the set $\{ \lm_{\Au} \}$, as a consequence we can construct an orthonormal frame in the background space at each point of $\CP^n$ as the columns of the adjoint representation of $SU(n+1)$:

\be \label{xiab}
\xi_a^b := -\half \tr(g \lm_b g^{\dagger} \lm_a)  = -Ad_{ba}(g), \qquad \xi^a_c \xi^b_c = \dl^{ab}~.
\ee
In this notation we have $\xi_a = \xi^{n^2+2n}_a $, which together with $\xi^{I}_a$ are normal, and $\xi^{A}_a$ which are vectors tangent to $\CP^n$. The left and right action differential operators which generate the Lie algebra $su(n+1)$ are given by \cite{Eguchi}

\be
\L_a = -\tr \Big( \frac{\lm_a}{2} g \frac{\partial}{\partial g^T} \Big) = -\frac{ (\lm_a)^\al_\bt}{2} \left(  z^\bt \del_\al  +  u^\bt_{\imath} \del^{~\! \!\imath}_\al \right), \quad 
\K_a = \tr \Big( g \frac{\lm_a}{2} \frac{\del}{\del g^T} \Big)~.
\ee
We remark that left actions are a moving frame rotation of right actions:

\be \label{LeftRight}
Ad_{ab}(g)\L_b [g] = - g \frac{\lm_a}{2}  = - \K_a [g], \qquad ~~\mbox{hence} ~~\qquad \xi^a_b  \L_b = \K_a ~.
\ee
In \cite{Huet} we proposed an ansatz\footnote{Such ansatz was shown to correspond to the canonical spin$_c$ Dirac operator, details may be found in \cite{Huet}.}
  for the Dirac operator on $\CP^2$ and to this end introduced the Clifford representation as the Lie algebra homomorphism $ su(n+1)\hookrightarrow spin(n^2 + 2n)$, given by:

\be \label{CliffordRepn}
T_a = \frac{1}{4i} f_{abc}\gm^{bc}, \qquad \gm^{bc} = \half [\gm^b, \gm^c],
\ee
through the exponential map we obtain a representation of $SU(n+1)$, which we will call $Cliff$; the homomorphic image of $g$ in $Cliff$ is denoted by $\g$. Here $\gm^a$ are the Clifford algebra generators for the Euclidean background space, they can be thought of as Dirac matrices satisfying 

\be
\{ \gm^a , \gm^b \} = 2 \dl_{ab} {\bf 1}.
\ee
The generators $T_a$ satisfy the same commutation relations as $\lm_a /2$. Notice that the defining relations of the left and right action operators translate to the Clifford 
representation as 

\be \label{Ad}
\L_a [\g] = -T_a \g , \qquad \K_a[\g] = \g T_a .
\ee
The Dirac matrices $\gm^a$ carry the adjoint representation of $SU(n+1)$ naturally since

\be \label{Adgm}
\g \gm^a \g^\dagger = Ad_{ab}(g)\gm^b,
\ee
also, $\CP^n$ is a symmetric coset space \cite{Nomizu}, therefore the structure constants satisfy the following conditions

\be \label{f0}
f_{\al \bt i} = 0, \quad f_{ijk} = 0~.
\ee
The first identity states that $s(u(n)\times u(1))$ is a subalgebra while the second states that the canonically induced $su(n+1)$ connection has vanishing torsion. Another useful fact is  the total antisymmetry of the structure constants $f_{abc}$ due to the compactness of $su(n+1)$ \cite{ORaif}. In our conventions the antisymmetrisation bracket denotes antisymmetrisation with an additional prefactor: $X_{[a_1 \cdots a_m]} := \frac{1}{m!} \sum_{\ep} \mbox{sgn} (\ep) X_{a_{\ep(1)} \cdots a_{\ep(m)}}$, e.g. $\gm^{a_1 \cdots a_m} := \gm^{[a_1} \cdots \gm^{a_m]}$.

\section{Dirac operator}

In \cite{Huet} an ansatz for the left acting Dirac operator on $\CP^2$ was proposed, there it was shown that such ansatz corresponds to the canonical spin$_c$ structure on $\CP^2$. We will show that the ansatz is actually a moving frame rotation of the standard Dirac operator constructed from right actions \cite{Fried}, and that such ansatz can be generalised to a universal operator on $\CP^n$ for all spin$_c$ bundles by choosing the domain suitably. In the general case both the domain and the spectrum depend on the chosen spin$_c$ structure \cite{Baer2}.

As our left acting Dirac operator we propose a natural extension of the operator presented in \cite{Huet}

\be \label{ansz}
\D = \gm^a P_{ab} (\L_b + T_b)  = \gm^a D_a , \qquad D_a := P_{ab}(\L_b  + T_b)~.
\ee
It differs in form from that presented in \cite{Huet} in that the domain of $\L_b$ has been extended to all differentiable functions over $SU(n+1)$, and not just on $ S(U(n)\times U(1))$-invariant functions. This means in particular that the left action can no longer be considered tangent as in \cite{CPNF} because $P_{ab}\L_b \neq \L_a$. We aim to show that this operator is universal in the sense that it retains its form for different spin$_c$ bundles, changing only its domain.

A universal Dirac operator based on right acting operators was constructed in \cite{Universal}, this operator is $\D_K := \gm^A \K_A$, we show first that our ansatz (\ref{ansz}) is, up to a sign, a moving frame rotation of $\D_K$. Dirac operators constructed from right actions are standard in the literature and may be found for coset spaces and symmetric riemannian manifolds in \cite{Bal,Fried}, but left acting Dirac operators are not so well known, the following proposition shows a connection between them. 

\newtheorem{L1}{Proposition}

\begin{L1} \label{p1}
Right and left action Dirac operators are related through 
\[
\D = -\g \circ \D_K \circ \g^{\dagger}
\]
\end{L1}

\vspace{.2cm}

{\it Proof:}
Consider the following series of transformations, notice that\footnote{\label{f2} This can be seen from the completeness relation $\xi_a^A \xi_b^A + \xi_a^{\Au} \xi_b^{\Au} = \delta_{ab}$ which restates the orthogonality of $Ad$ or by direct calculation, cf. footnote \ref{f4} .} $P_{ab} = \xi^A_a \xi^A_b$, and $Ad$ is orthogonal:

\bear
\gm^A \K_A &=& - \gm^A Ad_{Aa}(g)\L_a =  \gm^A \xi^A_a \L_a = \gm^A \xi^A_a P_{ab}\L_b = \gm^c \xi^c_a P_{ab}\L_b \nn \\
&=& -\gm^c Ad_{ca}(g)P_{ab}\L_b  = - Ad_{ac}(g^\dagger)\gamma^c P_{ab}\L_b = -\g^{\dagger} \gm^a \g P_{ab} \L_b  \\
&=& -\g^{\dagger} \circ (\gm^a P_{ab} \L_b - \gm^a P_{ab} \L_b[\g] \g^\dagger   ) \circ \g = - \g^\dagger \circ \gm^a P_{ab}(\L_b + T_b) \circ \g \qquad \blacksquare \nn
\ear

Proposition \ref{p1} explains why our ansatz worked in \cite{Huet} in the first place, and clarifies its hitherto unjustified origin. Before proceeding further we should note that the coordinates $\xi_a^b$ carry the adjoint representation with the index $a$ transforming under left actions and the index $b$ under right actions, as can be verified using (\ref{xiab}):

\be \label{Lxi}
\L_a [\xi_b^d] = if_{abc} \xi_c^d , \quad \K_a [\xi_b^d] = if_{adc}\xi^c_b.
\ee
It follows in particular from (\ref{Lxi}) that when restricted to act upon a differentiable function on $\CP^n$ one has $P_{ab} \L_b [F(\xi)] = \L_a [F(\xi)]$, as shown in \cite{CPNF}, so that in such case the operator $\L_a$ is indeed tangent.

We will construct the domain of the Dirac operator as the Hilbert space of all square integrable sections of the spin bundle with the correct representation content, we shall refer to such domain as the spinor space and denote it by $S_q$. A chirality operator will also be constructed allowing us to find the non-zero spectrum of the Dirac operator from the spectrum of its square

\be
\D^2 = D^2 + \frac{\gm^{ab} F_{ab}}{2}  + F_{\del}.
\ee
Above we have introduced the curvature of the covariant derivative $D_a$, $ F_{ab} = [D_a , D_b] - if_{abc} D_c$, the spin laplacian $D^2 = D_a D_a$, and defined the operator

\be \nn
F_{\del} = \frac{i}{2} \gm^{ab} f_{abc} D_c + \gm^a [D_a, \gm^b] D_b.
\ee
As a first step we show that this operator vanishes.

\newtheorem{L2}[L1]{Proposition}

\begin{L2} \label{p2}
The operator $F_{\del}$ vanishes identically, $F_{\del}=0$.
\end{L2}

\vspace{0.2cm}

{\it Proof:} Using (\ref{f0})

\bear
F_{\del} &=& \frac{i}{2} \gm^{ab} f_{abc} D_c + \gm^a [D_a, \gm^b] D_b = i\gm^{ae}(\half f_{aec} + P_{ab} f_{bce})D_c \nn \\
&=& \frac{i}{2} \gm^{ae}f_{aej}D_j + i \gm^{ie}f_{ije}D_j = i(\gm^{\ald i} + \gm^{i \ald} )f_{\ald ij}D_j = 0 ~~~\blacksquare
\ear

The curvature tensor found, $F_{ab}$, has an algebraic and a differential component:

\bear
F_{ab} &=& \mathcal{F}_{ab} + iP_{bb'}P^{\perp}_{ca'}f_{b'ac} \L_{a'}
\ear
the differential contribution vanishes in the case of canonical spin$_c$ structure because in such case the left actions $\L_a$ are tangent when acting on spinors, which are identified with differential forms \cite{Smilga}. We have introduced the complementary normal projector $P^{\perp}_{ab} = \delta_{ab} - P_{ab}$ above, and although it is not manifest the curvature tensor obtained is antisymmetric. The algebraic part $\mathcal{F}_{ab}$ is the curvature tensor of the canonical spin$_c$ structure given in \cite{Huet}, namely

\be
\mathcal{F}_{ab} = i( P_{ad}P_{be}f_{dec} + f_{abe}P_{ec} + f_{ace}P_{be} -f_{bce}P_{ae})T_c
\ee
it was shown there that $\frac{\gm^{ab}\mathcal{F}_{ab}}{2}= \frac{\gm^{ij} f_{ij \ald} } {2i} T_{\ald}$. Hence we can rewrite the curvature term

\be \label{R}
\frac{\gamma^{ab} F_{ab}}{2} = \frac{\gm^{ij}}{2i} f_{i j \ald }(\L_{\ald} + T_{\ald}) ~.
\ee 
It may seem at this point that $F_{ab}$ cannot be a curvature tensor as it contains a differential operator, however we will show later that when restricted to act on sections of the spin bundle the curvature is actually algebraic. Before proving that statement we will present some results needed for the computation of the spectrum in appendix \ref{B}, the spectrum found through this approach conforms to known results \cite{Cahen,Seifarth,Universal}.

\newtheorem{L3}[L1]{Proposition}

\begin{L3} \label{p3}
$D^2 \circ \g = \g \circ \K_A^2$ .
\end{L3}

\vspace{0.2cm}

{\it Proof:} 
The action of $\L_a$ on $\g$ leads to $ D^2 \circ \g = \g  P_{ab}\L_b \circ P_{ac}\L_c$, then since $ P_{ab} = \xi^A_a \xi^A_b$ along with (\ref{xiab}) one gets $D^2 \circ \g = \g \circ \xi^A_a \K_A[\xi^B_a] \K_B + \g \circ \K_A^2$, observe then from (\ref{Lxi}) $\K_A[\xi^B_a] = if_{ABe}\xi^e_a = if_{AB\Cu } \xi^{\Cu}_a$, then 
$\xi^A_a \K_A[\xi^B_a] \K_B = if_{AB\Cu} \delta_{\Cu A} \K_B = 0$ $\blacksquare$

\newtheorem{L4}[L1]{Proposition}

\begin{L4} \label{p4}
$ \gm^{ij}f_{ij\ald} (\L_{\ald} + T_{\ald}) \circ \g = - \g \circ \gm^{AB}f_{AB \Cu} \K_{\Cu}$ .
\end{L4}

\vspace{0.2cm}

{\it Proof:} Observe that 

\be
\gm^{ij}f_{ij \ald} (\L_{\ald} + T_{\ald}) \circ \g = \gm^{ij}f_{ij \ald} \g \circ \L_{\ald} = \gm^{ij}f_{ij a} \xi^{\Cu}_a \g \circ \xi^{\Cu}_{b} \L_{b} = \g \circ (\gm^{ij}f_{ija}\xi^{\Cu}_a  )^0 \K_{\Cu}  
\ee 
We need to show first that $\gm^{ij} \circ f_{ij a} \xi^{\Cu}_a \g  = \g \circ (\gm^{ij}f_{ij a} \xi^{\Cu}_a)^0$, where the superindex $0$ means that the expression is evaluated at the north pole, to this end we rewrite $\gm^{ij} \circ f_{ij a} \xi^{\Cu}_a \g  = \gm^{ab} \circ P_{aa'}P_{bb'} f_{a'b'c'} \xi^{\Cu}_{c'} \g$ and use the $su(3)$ invariance of the structure constants, the property (\ref{Adgm}) and the orthogonality of $Ad$. Finally observe that evaluating the expression at the north pole through $(\xi^e_c)^0 = Ad(g)_{ca} \xi^e_a = - \dl^e_c$ yields $\g \circ (\gm^{ij}f_{ij a} \xi^{\Cu}_a)^0 = -\g \circ \gm^{AB} f_{AB \Cu}$ $\blacksquare$

We define now the following north pole operators
\be
\phi^0 := \frac{i \gm^{AB} f_{AB(n^2+2n)}}{\sqrt{2 n(n+1)} }, \qquad \tau^0_{I} := \frac{\gm^{AB}f_{ABI}}{4i}~,
\ee
and their transported values: $\phi = \g \phi^0 \g^{\dagger} =-\frac{i}{2n} \gm^{ab}J_{ab} $, $\tau_I = \g \tau^0_I \g^{\dagger}$. The operator $\phi$ we have defined generalises the hypercharge for $\CP^2$ wich was introduced in \cite{Huet}, and together with the operators $\tau_I$ furnish a point-dependent representation of the isotropy algebra $s(u(n)\times u(1))$ within the Clifford representation. We can now restate proposition \ref{p4} as the formula

\be \label{curvident}
\frac{\gm^{ab}F_{ab}}{2} \g = -2\g \tau_I^0 \K_I + \g \phi^0 \K_0~,
\ee
where we introduced, for later convenience, $ \K_0 := \sqrt{\frac{n(n+1)}{2}} \K_{n^2 + 2n}  =  \half (u_{\imath}^\al \del^{\imath}_{\al} - n z^{\al}\del_{\al})$.

\section{Clifford algebra and spinor space} \label{cliffandspin}

It is convenient to introduce a complex basis of the Clifford algebra of the background euclidean space $\mathbb{R}^{n^2 +2n}$, this can be donde by pairing the Dirac matrices appropriately. The pairing is naturally done into holomorphic combinations when $n$ is even, and when $n$ is odd we introduce an extra gamma matrix $\gm^{n^2 + 2n + 1}$ to complete the pairings at the cost of doubling the dimension of the Dirac matrices. The holomorphic basis can be constructed as follows:

\be
\Gamma^{\imath} := \frac{\gm^{n^2 + 2 \imath - 2}  - i \gm^{n^2 + 2\imath - 1}}{2} ~.
\ee
In our notation we will use for the values $\J = 1,2, \cdots, n$ the indices $\imath, \jmath$ instead of $\I,\J$ unless otherwise indicated. There are two cases, when $n$ is even or odd, in the case when $n$ is even we define the remaining pairs as

\be
\Gamma^{\J} = \frac{\gm^{2s -1}  - i \gm^{2s}  }{2}, \qquad \mbox{for} \qquad s= 1, 2, \cdots, \frac{n^2}{2}-1 ~,
\ee
where $ \J = n+s$, and the last two remaining matrices are paired into $\Gamma^{n+ \frac{n^2}{2}} = \half(\gm^{n^2-1} - i \gm^{n^2+2n})$. If $n$ is odd we need to introduce an extra gamma matrix $\gm^{(n+1)^2}$, change the range of $s$ to $s= 1, \cdots, (n^2-1)/2$ and pair the last two gamma matrices into $\Gamma^{(n+1)^2/2} = \half (\gm^{n^2 + 2n} - i \gm^{(n+1)^2})$. 

The complex basis of matrices includes also the anti-holomorphic matrices $\Gamma^{\bar{\J}} = (\Gamma^{\J})^{\dagger}$ and satisfies the anticommutation relations:  

\be
\{ \Gamma^{\I}, \Gamma^{\J} \}  = \{ \Gamma^{\bar{\I}}, \Gamma^{\bar{\J}} \}  =  0 , \quad \{ \Gamma^{\I} , \Gamma^{\bar{\J}} \} = \dl^{\I}_{\J} , \qquad \I,\J =1,\cdots, 
n+ \lceil n^2/2 \rceil~.
\ee
We now introduce a Clifford vacuum state, $\o$, as the state that is annihilated by all holomorphic gamma matrices: $\Gamma^{\J} \o =0$. With these conventions, an analysis of the structure constants yields the following useful expressions:

\be
\tau_I^0 = (\frac{\lm_I}{2})^{\jmath}_{\imath} \Gamma^{\bar{\jmath}\imath}, \qquad \phi^0 = \frac{2}{n} \Gamma^{\imath \bar{\imath}}  \quad \mbox{where} \quad \Gamma^{\I \bar{\J}} = \half [\Gamma^{\I}, \Gamma^{\bar{\J}}]~,
\ee
and the identities

\bear
\left[ \phi^0, \Gamma^{\bar{\imath}} \right] = -\frac{2}{n} \Gamma^{\bar{\imath}},& & \qquad \left[ \phi^0, \Gamma^{\imath} \right] =  \frac{2}{n} \Gamma^{\imath}, \qquad \qquad~ 
\phi^0 \o = \o , \nn \\ 
\left[ \tau_I^0, \Gamma^{\bar{\imath}} \right] = (\frac{\lm_I}{2})^{\jmath}_{\imath} \Gamma^{\bar{\jmath}},& & \qquad \left[ \tau_I^0, \Gamma^{\imath} \right]=
 - (\frac{\lm_I}{2})^{\imath}_{\jmath} \Gamma^{\jmath}, \qquad \tau^0_I \o = 0 ~.
\ear
A spinor field on $\CP^n$ is a section of the spin$_c$ bundle, the space of such sections can be constructed from the Dolbeault complex \cite{Universal,Smilga}:

\be
S_q (\CP^n) =   \bigoplus_{k=0}^{n} \Omega^{(0,k)}( \CP^n ) \otimes \mathfrak{L}^{q} ~.
\ee
Spinor fields can then be conceived as linear combinations of charged anti-holomorphic forms of any degree, $\mathfrak{L}$ above is the tautological line bundle. An arbitrary spinor field decomposes into \cite{GSW}

\be
\Psi = \Psi^0 \o + \Psi^{\imath} \Gamma^{\bar{\imath}} \o + \Psi^{\imath \jmath} \Gamma^{\bar{\imath}}\Gamma^{\bar{\jmath}} \o + \cdots + \Psi^{12\cdots n}\Gamma^{\bar{1}} \cdots \Gamma^{\bar{n}} \o ~.
\ee
The eigenspinors of the non-zero spectrum of $\D^2$ can be written taking into account the representation content of the $\CP^n$ spin$_c$ bundle given in \cite{Cahen,Universal}

\be \label{eigenspin}
\Psi^{(\sigma)} = 2^k ~\!\nu_k(l,q,n) \zb_{\al_{k+1}} \cdots \zb_{[\al_{l+q}} \bar{u}^{\imath_1}_{\al_1} \cdots \bar{u}^{\imath_k}_{\al_k ]} z^{\bt_1} \cdots z^{\bt_l} \g \Gamma^{\bar{\imath}_1} \cdots \Gamma^{\bar{\imath}_k} \o~.
\ee
The factor $2^k \nu_k(l,q,n)$ is a convenient normalisation factor which is calculated in appendix \ref{A} and $(\sigma)$ stands for the multiple label $(l,q,n,k)$. After manipulating the expression above one may show that the eigenspinors can also be rewritten as 

\be \label{eigenspin2}
\Psi^{(\sigma)} = \nu_k(l,q,n) \zb_{\gm_1} \cdots \zb_{\gm_k} \zb_{\al_{k+1}} \cdots \zb_{[\al_{l+q}}(\lm_{a_1})^{\gm_1}_{\al_1} \cdots (\lm_{a_k})^{\gm_k}_{\al_k ]}
 z^{\bt_1} \cdots z^{\bt_l} \gm^{a_1 \cdots a_k} \om~,
\ee
where the antisymmetrisation bracket affects only greek indices. We remark that $\om = \g \o$, the {\it transported vacuum state}, can be identified as the zero mode in the canonical spin$_c$ structure as observed below. The zero modes of the Dirac operator span its kernel, which we call $K$. It is a known result \cite{Witten} that for a self adjoint operator $\ker(\D^2) = \ker(\D)$, and the representation content of the zero modes leads us to their explicit form:

\bear
 \chi &=& \nu z^{\al_1} \cdots z^{\al_{|q|}} \om~, \qquad q \leq 0~, \nn \\
 \psi &=& \nu \zb_{\al_1} \cdots \zb_{\al_{q-n-1}} \omb~, \qquad q \geq n+1~. 
\ear
Where we introduced $\omb = \g\Gamma^{\bar{1}} \cdots \Gamma^{\bar{n}}\o$ and the normalisation constant $\nu = 1/\sqrt{\mbox{Vol}(\CP^n)}$ found in appendix \ref{A}. These zero modes were first obtained in a different manner by Wipf et al. in \cite{Wipf} through a local coordinate analysis. The choice $q=0$ corresponds to the canonincal spin$_c$ structure \cite{Universal,Lawson}, in such case the spinors are neutral differential forms and the only zero mode is $\om$; we will find the explicit form of $\om$ in section \ref{zmodes}, where we also show that $\omb$ is in a sense conjugate to $\om$. In particular it follows from our approach that all eigenspinors can be determined when $\om$ alone is known.

With these considerations we are now able to construct the spinor space as

\be
S_q := \bigoplus_{ k=0,~ l}^{n-1} \langle \Psi^{(\sigma)} \oplus \D \Psi^{(\sigma)} \rangle  \oplus K~,
\ee
where the brackets denote the linear hull taken over the ring of smooth functions on $\CP^n$, $S_q$ is a projective submodule of the module of sections of the trivial spin bundle associated to the background euclidean space, the smallest which is $\D$-invariant and contains $K$ and all the $\Psi^{(\sigma)}$. Spinor fields are defined as the elements of the spinor space and are always square integrable on account of the compactness of $\CP^n$. The non zero eigenspinors of the Dirac operator are constructed in the standard manner as:

\be
\Psi^{(\sigma)}_{\pm} =  \Psi^{(\sigma)} \pm  \frac{\D \Psi^{(\sigma)} }{\sqrt{\Lambda_{\sigma}}}~.
\ee
where $\Lambda_{\sigma}>0$ is the eigenvalue of $\D^2$. We now prove that the curvature tensor (\ref{R}) is algebraic, as opposed to differential, when restricted to the spin bundle $S_q$.

\newtheorem{T1}{Theorem}

\begin{T1} \label{t1}
The operators $\xi^{\Au} \cdot (\L + T)$ are algebraic in $S_q$, in fact:

\be
\xi \cdot (\L + T)|_{S_q} = \sqrt{\frac{n}{2(n+1)}} \left( q + \frac{(n+1)}{2}(\phi -1) \right)|_{S_q}, \quad \xi^I \cdot (\L + T)|_{S_q} = -\tau_I |_{S_q}.
\ee
\end{T1}

{\it Proof} First we prove that they coincide in each subspace $\langle \Psi^{(\sigma)} \rangle$ and in $K$. It is easiest to transport to the north pole and use equivariance. Notice $\xi \cdot (\L + T) [\Psi^{(\sigma)}] =  \frac{\sqrt{2}}{\sqrt{n(n+1)}} \g \K_0 [\g^{\dagger} \Psi^{(\sigma)}] $, a simple calculation leads to 

\be
\xi \cdot (\L + T)[\Psi^{(\sigma)}] = \frac{nq -k(n+1)}{\sqrt{2n(n+1)}} \Psi^{(\sigma)},
\ee
now notice that 

\be
\phi \Psi^{(\sigma)} = \g \phi^0 \g^{\dagger} \Psi^{(\sigma)} = \Big(\frac{n-2k}{n}\Big)\Psi^{(\sigma)}~.
\ee
Upon substitution we can see that both operators coincide when acting on $\Psi^{(\sigma)}$. They also coincide on the kernel $K$, since

\bear
\xi \cdot (\L + T) [\chi] &=& q \sqrt{\frac{n}{2(n+1)}} \chi, \qquad \phi  \chi =  \chi~, \nn \\
\xi \cdot (\L +T)  [\psi] &=& \frac{(q-n-1)\sqrt{n}}{\sqrt{2(n+1)}}  \psi, \quad \phi \psi = - \psi~, \nn
\ear
from which the result follows on $K$. Next we concern ourselves with the second operator, observe that $\tau_I \Psi^{(\sigma)} = \g \tau_I^0 \g^{\dagger} \Psi^{(\sigma)}$ and that in fact $\tau_I^0 = \half (\lambda_I)^{\jmath}_{\imath} \Gamma^{\bar{\jmath}\imath}$, hence

\bear \label{tensorch}
\g \tau_I^0 \bar{u}^{\imath_1}_{\al_1} \cdots \bar{u}^{\imath_k}_{\al_k} \Gamma^{\bar{\imath}_1} \cdots \Gamma^{\bar{\imath}_k} \o &=& 
\g \bar{u}^{\imath_1}_{\al_1} \cdots \bar{u}^{\imath_k}_{\al_k} [ \tau_I^0, \Gamma^{\bar{\imath}_1} \cdots \Gamma^{\bar{\imath}_k}] \o  \\
&=& \g \sum_{p=1}^k \bar{u}^{\imath_1}_{\al_1} \cdots \bar{u}^{\imath_k}_{\al_k} \Gamma^{\bar{\imath}_1}\cdots \frac{(\lm_I)^{\jmath}_{\imath_p}}{2}\Gamma^{\bar{\jmath}} \cdots 
\Gamma^{\bar{\imath}_k} \o \nn \\
&=& - \g \K_I [\bar{u}^{\imath_1}_{\al_1} \cdots \bar{u}^{\imath_k}_{\al_k} \Gamma^{\bar{\imath}_1} \cdots \Gamma^{\bar{\imath}_k} \o] \nn \\
&=& - \xi^I \cdot (\L + T) \g \bar{u}^{\imath_1}_{\al_1} \cdots \bar{u}^{\imath_k}_{\al_k} \Gamma^{\bar{\imath}_1} \cdots \Gamma^{\bar{\imath}_k} \o ~.\nn
\ear
The action also coincides in the kernel, because both $\xi^{I} \cdot (\L + T)$ and $\tau_I$ vanish since the isotropy subalgebra generators satisfy $\K_I[z] = \K_I[\bar{z}]=0$,

\bear
\xi^{I} \cdot (\L + T) \chi = \g \K_I [\g^{\dagger} \chi] = 0, \quad \tau_I  \chi = \g \tau_I^0 \g^{\dagger} \chi  = 0 ~,\nn \\
\xi^{I} \cdot (\L + T) \psi = \g \K_I [\g^{\dagger} \psi] = 0, \quad \tau_I \psi = \g \tau_I^0 \g^{\dagger} \psi  = 0 ~,\nn 
\ear
where the last line follows from

\bear
\tau_I^0 \Gamma^{\bar{1}} \cdots \Gamma^{\bar{n}} \o &=& [\tau_I^0, \Gamma^{\bar{1}} \cdots \Gamma^{\bar{n}}] \o = 
\sum_{p=1}^{n} \Gamma^{\bar{1}} \cdots \frac{(\lm_I)^{p}_{\jmath}}{2} \Gamma^{\bar{\jmath}} \cdots \Gamma^{\bar{n}} \o  \nn \\
&=& \sum_{p=1}^{n} \Gamma^{\bar{1}} \cdots \frac{(\lm_I)^{p}_{p}}{2} \Gamma^{\bar{p}} \cdots \Gamma^{\bar{n}} \o = \half \tr(\lm_I) \Gamma^{\bar{1}} \cdots \Gamma^{\bar{n}} \o =0 ~.\nn
\ear
We prove next that the operators coincide in the last subspace, generated by $\langle \D  \Psi^{(\sigma)} \rangle$. To this end observe 
that since $\D \circ \g = -\g \circ \D_K$ and we may write $\D_K = \Gamma^{\imath} \K_{\imath} + \Gamma^{\bar{\imath}}\K_{\bar{\imath}}$ the action of, say $ \Gamma^{\imath} \K_{\imath}$, (the same occurs with $\Gamma^{\bar{\imath}}\K_{\bar{\imath}}$) upon $\g^{\dagger}\Psi^{(\sigma)}$ has the following property: $\K_{\imath}$ increases its eigenvalue associated to $\K_0$ by $\frac{n+1}{2}$ whilst $\Gamma^{\imath}$ increases the eigenvalue of $\phi^0$ by $\frac{2}{n}$, producing thus an increment on the eigenvalue of $\frac{n(n+1)}{4}(\phi^0 -1)$ of the amount 
$\frac{n(n+1)}{4} \frac{2}{n} = \frac{n+1}{2}$, hence the operators coincide because they do on $\g^{\dagger}\Psi^{(\sigma)}$, and this was already established. Finally, we notice that $\D_K$ is an $su(n)$ scalar, it cannot change the tensor structure of $\g^{\dagger}\Psi^{(\sigma)}$, exhibited in (\ref{tensorch}), but will produce linear combinations of terms each with the same structure, hence the identity will remain valid on $ \langle \D \Psi^{(\sigma)} \rangle ~\blacksquare$

Theorem \ref{t1} allows us to write expressions for the curvature of the spin$_c$ bundle and the Dirac operator in terms of left actions, which exhibit explicitely the spin connection and its dependance upon the spin$_c$ structure:

\be
\D = \gm_a \left( \L_a + T_a -  \frac{\sqrt{n}\xi_a}{ \sqrt{8(n+1)}}[2q +(n+1)(\phi-1)] + \xi^I_a \tau_I \right)
\ee
and

\be \label{curvature}
F_{ab} = \mathcal{F}_{ab} - iP_{aa'} P_{bb'} f_{a' b' c'} \left( \frac{\sqrt{n}\xi_{c'}}{\sqrt{8(n+1)}} (2q-n-1)  - \xi_{c'}^I (\tau_I + \xi^I \cdot T) \right)~.
\ee
As one should expect these expressions are given completely in terms of representation theory, and show that the additional $U(1)$ part of the holonomy in the curvature tensor vanishes exactly for the unique spin structure choice $q=\frac{n+1}{2}$. It is also patent in these identities, valid on the spinor space, that the curvature is algebraic, as stated before.

\section{Chirality}

Several equivalent definitions of the chirality can be given, which, up to a phase, is the product of all tangent gamma matrices at each point of the manifold. We define the chirality operator as

\be
\Gamma := i^n \g \prod_{A} \gm^A \g^\dagger = \g \Gamma^0 \g^{\dagger}, \qquad \Gamma^2 = {\bf 1}~,
\ee
a direct calculation yields

\be \label{chirality}
\Gamma = \g \prod_{\imath = 1}^n [\Gamma^{\imath},\Gamma^{\bar{\imath}}] \g^\dagger, \quad \{\Gamma^0 , \Gamma^{\imath} \} =\{\Gamma^0 , \Gamma^{\bar{\imath}} \}=0~.
\ee
It is also possible to express the chirality in terms of the complex structure, to this end let $[Y]_{max}$ denote the projection of $Y$ onto the highest degree component of the graded Clifford algebra, and observe that

\bear
[\phi^n]_{max} &=& \frac{n!}{n^n}  \g \prod_{\imath = 1}^n [\Gamma^{\imath} , \Gamma^{\bar{\imath}}]  \g^{\dagger} =
\left(\frac{-i}{2n} \right)^n [J_{a_1 b_1} \cdots J_{a_n b_n} \gm^{a_1 b_1} \cdots\gm^{a_n b_n} ]_{max}   \nn \\
&=& \left(\frac{-i}{2n} \right)^n J_{a_1 b_1} \cdots J_{a_n b_n} \gm^{a_1 b_1 \cdots a_n b_n} 
\ear
leads to the alternative expression

\be \label{GJ}
\Gamma = \frac{(-i)^n}{2^n n!}J_{a_1 b_1} \cdots J_{a_n b_n} \gm^{a_1 b_1 \cdots a_n b_n} ~.
\ee
Even another expression is possible in terms of the tangent frame:

\be \label{Gxi}
\Gamma = i^n \prod_{A}(\g \gm^A \g^\dagger) = i^n \prod_{A}(\xi_a^A \gm^a) = i^n \xi_{a_1}^{n^2} \cdots \xi_{a_{2n}}^{n^2 +2n-1} \gm^{a_1 \cdots a_{2n}}~.
\ee
Through direct calculation one can prove that the chirality operator satisfies $\{ \D, \Gamma \} =0$.

Now we define the following operator, the chirality of the background space (which if necessary is extended to have even dimension, as described in section \ref{cliffandspin})

\be \label{Chirality}
\gm = \prod_{\J = 1}^{ n + \lceil n^2 /2 \rceil  } [\Gamma^{\J} ,\Gamma^{\bar{\J}}]   , \qquad \gm^2 = {\bf 1}, \quad \{ \D , \gm \} =0, \quad \{\gm, \gm^a \} =0 ~,
\ee
notice that $\gm$ is invariant under the adjoint action: $\g \gm \g^\dagger = \det(Ad(g))\gm = \gm $ and that $[\L_a + T_a ,\gm] =0$.

We are now in condition to prove a theorem, which states that both, the coordinate dependent chirality $\Gamma$ and the constant chirality $\gm$, agree on sections of the spin bundle:

\newtheorem{T2}[T1]{Theorem}

\begin{T2} \label{t2}
The restrictions of both chirality operators to the spinor space are identical: $$\Gamma|_{S_q} = \gm|_{S_q}~.$$
\end{T2}

{\it Proof}: This is a consequence of $\Gamma \g \Gamma^{\bar{\imath}_1 } \cdots \Gamma^{\bar{\imath}_k }\o = \gm \g \Gamma^{\bar{\imath}_1 } \cdots \Gamma^{\bar{\imath}_k }\o $,
a simple manipulation shows that they will coincide when $\Gamma^0 \o = \gm \o$, that this is indeed the case is easy to verify by direct calculation using (\ref{chirality}) and (\ref{Chirality}), in fact one has the stronger identity

\be
\gm \g \Gamma^{\bar{\imath}_1 } \cdots \Gamma^{\bar{\imath}_k }\o  = \Gamma \g \Gamma^{\bar{\imath}_1 } \cdots \Gamma^{\bar{\imath}_k }\o = 
(-1)^k \g \Gamma^{\bar{\imath}_1 } \cdots \Gamma^{\bar{\imath}_k }\o 
\ee
so that $\om$ has actually even chirality and the filled state $\omb$ has $(-1)^{n}$ chirality $\blacksquare$

This result shows that we can use $\gm$ as the chirality, where all dependance on the coordinates has disappeared, as a simpler but equivalent form of the chirality.

\section{Structure of the zero modes $\om$, $\omb$ } \label{zmodes}

In this section we discuss the essential structure of the canonical spin$_c$ structure zero mode $\om$, since all eigenspinors can be found if we know $\om$. We determine $\om$ completely using group-theoretical arguments. First notice that both $\om, \omb$ are $SU(n+1)$ scalars because $\om, \omb \in \bigcap_{a} \ker (\L_a + T_a)$, and hence using $  \xi \cdot (\L + T) \om =0$ along with $\xi \cdot \L = \frac{\sqrt{2}}{\sqrt{n(n+1)}} \K_0$ and $ \xi \cdot T = \phi \sqrt{n(n+1)/8}$ it follows that

\be \label{phiom}
\phi \om = \om  = -\frac{4}{n(n+1)} \K_0 \om
\ee
which implies 

\be \label{Komega}
\K_0 \om = -\frac{n(n+1)}{4}\om~.
\ee
We show in appendix \ref{C} that the Clifford representation is a sum of copies of the irreducible representation\footnote{We label representations through their Dynkin indices vector.} $R = (1,1, \cdots, 1)$, i.e. $Cliff = \bigoplus_r R$. Since $\om$ is a scalar this means that we must 
couple the bundle coordinates for $g$ with the representation $R$ to form an $SU(n+1)$ scalar, and this is only possible if we couple a representation $\bar{R} = R$. Such representation will consist of a tensor product of $\rho = \frac{n(n+1)}{2}$ bundle coordinates. The only possibility to satisfy (\ref{Komega}) is when each bundle coordinate is a $\bar{u}^{\imath}_{\al}$, as each will have a charge $-\half$. Therefore we have concluded that the zero mode is, up to a phase, a linear superposition of vectors of the form:

\be
\bar{u}^{\imath_1}_{\al_1} \cdots \bar{u}^{\imath_{\rho}}_{\al_{\rho}} e^{\al_1 \cdots \al_{\rho}}~.
\ee
Here $ e^{\al_1 \cdots \al_{\rho}}$ is the unitary basis of the tensor product representation $R$, it is a projection of the tensor product of $\rho$ basis vectors $e^{\al}$ of the 
fundamental representation. Now we observe that at the north pole $\g = \mathfrak{e}$ and $\om = \o$, since at the north pole we 
have $(\bar{u}^{\imath}_\al)^0 = \dl^{\imath}_\al$ this means that the vacuum $\o$ must be a superposition of the vectors $e^{\imath_1 \cdots \imath_\rho}$.

After the main result in appendix \ref{C} it becomes clear that such linear superposition consists of a single vector in each copy, the highest weight vector of the representation $R$ (since it is unique). This also determines the set $\{ \imath_1 , \imath_2 , \cdots , \imath_\rho \}$ uniquely and, fixing the phase properly, establishes that the vacuum state 

\be
\o = e^{n \cdots 2 2 1}
\ee
is the tensor corresponding to the highest weight filling of the Young tableau of $R$:

{\LARGE$$
\mbox{ \raisebox{0.34cm}{  $\young(\n \Cdots \Cdots\n)$    }  } \hspace{-3.075cm}
\mbox{\raisebox{-.34cm}{$\young(\nmo\Cdots) \hspace{-.1mm} \young(\nmo)$}}  \hspace{-2.036cm} 
\mbox{\raisebox{-1.35cm}{$\young(\Cdots\Cdots,\one)$}}
$$}
Taking these facts together we have found the following explicit form for the canonical zero mode:

\be
\om= \bar{u}^{n}_{\al_1} \cdots \bar{u}^{1}_{\al_{\rho}} e^{\al_1 \cdots \al_{\rho}}
\ee
Similar considerations lead to $\phi \omb = - \omb$, $\K_0 \omb = - \frac{n(n+1)}{4} \omb$ and the structure $\omb = u^{\bt_1}_{\jmath_1} \cdots u^{\bt_\rho}_{\jmath_\rho} e_{\bt_1 \cdots \bt_\rho}$, where the tensor $e_{\bt_1 \cdots \bt_\rho}$ corresponds to the lowest weight filling of $R$; these two tensors are dual to each other in a sense, one can see that $\omb$ is the transported highest weight state and is the state conjugated, under the Clifford algebra complexification, to $\om$. The state $\omb$ has associated the Young diagram with the lowest weight filling, which is constructed by filling the rows in the inverse order, placing $1$ inside the top row and decreasing one unit in each row so that $n$ fills the lowest box.

\section{Extended SUSY}

We show how the extended $\mathcal{N} =2$ supersymmetry is realised by means of a superhamiltonian $H$ and its two square roots, one of which is the Dirac operator. The square of the Dirac operator, $\D^2 = H = H^{\dagger}$, serves as the superhamiltonian, we identify two real supercharges:

\be \label{susyQ}
Q_1 = \frac{\D}{\sqrt{2}}, \quad Q_2 := \frac{ \gm^a \bar{D}_a}{\sqrt{2}}~,
\ee
with $\bar{D}_a := J_{ab}(\L_b + T_b)$. This superhamiltonian describes a non-relativistic fermion moving on $\CP^n$ in a static background field \cite{Brian}. We show now that the supercharges satisfy the superalgebra relations:

\be \label{susy}
\{Q_i ,Q_j \} = \delta_{ij} H, \qquad i,j =1,2.
\ee
To this end we carry out a short calculation:

\be
2(Q_2)^2 = \bar{D}^2 + \gm^{ab}\left( \half [\bar{D}_a, \bar{D}_b] - iJ_{ac}f_{cbe}\bar{D}_e \right)~.
\ee
where $\bar{D}^2 = \bar{D}_a \bar{D}_a$, and compare against

\be
H = D^2 + \gm^{ab} \left( \half [D_a, D_b] - iP_{ac}f_{cbe}D_e \right).
\ee
A straightforward calculation taking into account (\ref{f0}) shows at once that $\bar{D}^2 =D^2$. The remaining part of $2(Q_2)^2$ equals also that of $\D^2$, this fact follows from the next proposition, after some calculations:

\newtheorem{L5}[L1]{Proposition}

\begin{L5} \label{p5}
The identities $P_{ad}P_{bf} f_{def} = J_{ad}J_{bf} f_{def}$ and $P_{ac}J_{ef}f_{ceb} = J_{ac}P_{ef}f_{ceb}$ hold.
\end{L5}

\vspace{0.2cm}

{\it Proof:} Using Jacobi's identity and the condition of zero torsion (\ref{f0}) it follows that

\bear
P_{ad}P_{bf}f_{def} &=& J_{ag}J_{dg}J_{bc}J_{fc}f_{def} = J_{ag}P_{bf}J_{de}f_{dgf} + J_{ag}J_{bc}J_{fc}J_{df}f_{deg} \nn \\
&=& -J_{ag}J_{bc}P_{dc}f_{deg} = J_{ad}J_{bf}f_{def} \nn 
\ear
for the second identity contract the first identity with $J_{cb}$ and rename indices. $\blacksquare$

It is also useful to define the complex supercharges:

\be
Q = \frac{1}{\sqrt{2}}(Q_1 + iQ_2) = \gm^a K^+_{ab} J_b, \quad Q^{\dagger} = \frac{1}{\sqrt{2}}(Q_1 - iQ_2) = \gm^a K^{-}_{ab}J_b
\ee
where $J_a = \L_a + T_a$. 

\newtheorem{L6}[L1]{Proposition}

\begin{L6}
The supercharges $Q_i$ satisfy the superalgebra relations $\{ Q_i, Q_j \} = \delta_{ij} H$.
\end{L6}

\vspace{0.2cm}

{\it Proof:} At this point we need only show that $Q^2 =0$, which already implies the remaining relation $\{ Q_1, Q_2\} =0$. To this end observe that proposition \ref{p5} leads straightforwardly to the identity $K^+_{ac}K^+_{bd}f_{cde} = 0$ and notice also that $K^+_{ab}K^+_{ac} = 0$ since $K_{ab}^+ = K_{ba}^-$ and these are orthogonal projections. Therefore 

\be
Q^2 = \gm^a \gm^b K_{ac}^{+} K_{bd}^{+} J_{c}J_{d} + i \gm^{a}\gm^{b}K_{ac}^{+}K_{be}^{+}f_{cde}J_d = 0
\ee
and the superalgebra relations (\ref{susy}) follow. $\blacksquare$

\subsection{R-symmetry and superpotential}

Following \cite{Wipf} it is possible to define fermionic creation and annihilation operators\footnote{\label{f4} The assignation of creation and annihilation operators is inherited from the convention taken for complex gamma matrices and the identity $2 \xi^{\bar{\imath}}_a \xi^{\imath}_b  = K_{ab}^{+}$, cf. appendix \ref{B}}

\be
\psi_a := K^{+}_{ab} \gm^b,\quad \psi_a^\dagger = K^{-}_{ab} \gm^b
\ee
which satisfy the relations

\be
\{ \psi_a , \psi_b \} = \{\psi_a^\dagger , \psi_b^\dagger \} =0, \quad  \{\psi_a, \psi_b^\dagger \} = 2 K_{ab}^{+}~.
\ee
As pointed out also in \cite{Wipf}, the particle number operator is $N= \half \psi_a^{\dagger} \psi_a =  \frac{1}{2}(n + \frac{i}{2}J_{ab}\gamma^{ab})$, hence we may relate $\phi$ to the number operator: $N = \frac{n}{2}(1-\phi)$, since the number operator is the generator of the $R$-symmetry of the superalgebra we must conclude from here that the $R$-symmetry of the superalgebra is the $U(1)$ part of the holonomy of $\CP^n$, generated by $\phi$ (the full holonomy is generated by $\phi, \tau_I$). The number operator acts on the states generated by action of the creation operators on the {\it transported vacuum} while $\psi_a \om = \psi^{\dagger}_a \omb =0$.

It is also natural to split the Dirac operator into holomorphic and antiholomorphic contributions,

\be
\D = \psi_a \nabla_a^- + \psi_a^{\dagger}\nabla_a^+ = Q^{\dagger} + Q, \quad \quad \nabla_a^{\pm} = K^{\pm}_{ab}J_b 
\ee
and one finds that the fermionic creation and annihilation operators are covariantly constant in the sense $ [\nabla_a^+, \psi^{\dagger}_b] =[\nabla_a^-, \psi_b] = 0$. A superpotential $\Phi(g) \in Cliff(SU(n+1))$, for the spin$_c$ connection is defined by the property

\be \label{suppot}
\nabla_a^{\pm} = \Phi(g) \circ \L_a^{\pm} \circ \Phi(g)^{-1}, \quad \quad \L_a^{\pm} = K_{ab}^{\pm} \L_{b}.
\ee
We investigated the existence of such superpotential, our main result is proposition \ref{p7}, generalised in section \ref{lastsec}:

\newtheorem{L7}[L1]{Proposition}

\begin{L7} \label{p7}
The most general solution for the superpotential is $\Phi(g) = \g \mathfrak{K}$, with any fixed element $\mathfrak{K} \in Cliff(SU(n+1))$.
\end{L7}

\noindent The right multiplication by $\mathfrak{K}$ is naturally interpreted as a gauge freedom of the superpotential. In order to prove proposition \ref{p7} we need to establish some previous results. Before proceeding we will remark that $\Phi$ in a sense is not a true superpotential as the one found in \cite{Wipf}, this is because we are using a global coordinate system $\xi$ and the superpotential is a local section in a non trivial bundle. Nevertheless we found that it is possible in our approach to give a globally defined superpotential, but only if it is lifted to the $SU(n+1)$-bundle, as we shall see in what follows. To this end we use the following result

\newtheorem{L8}[L1]{Proposition}

\begin{L8} \label{p8}
 $\cap_a \ker(\L_a + T_a) = \mbox{im }\! \g$.
\end{L8}

\vspace{0.2cm}

{\it Proof:} We need only prove the inclusion $\cap_a \ker(\L_a + T_a) \subset \mbox{im }\! \g $. An element annihilated by all $\L_a + T_a$ must be an $SU(n+1)$ scalar, the most general such scalar has the form 

\be
\sum_{k} C^k_{\alpha_1 \cdots \alpha_M} u^{\alpha_1}_{\beta_1} \cdots u^{\alpha_M}_{\beta_M}e^{\beta_1 \cdots \beta_M}_k
\ee
where $C^k_{\alpha_1 \cdots \alpha_M} \in \mathbb{C}$, $M =\frac{n(n+1)}{2}$ and $1 \leq k \leq 2^{\lceil n/2\rceil}$ labels the copies of $R$ into which $Cliff$ decomposes (cf. appendix \ref{C}). Observe that $u^{\alpha_1}_{\beta_1} \cdots u^{\alpha_M}_{\beta_M}e^{\beta_1 \cdots \beta_M}_k = \g e^{\alpha_1 \cdots \alpha_M}_k ~\blacksquare$

To prove proposition \ref{p7} we rewrite the defining property (\ref{suppot}) as $ K_{ab}^{\pm} (\L_b + T_b)\Phi = 0$, which is equivalent to either of the following conditions

\be \label{equiv}
P_{ab}(\L_b + T_b) \Phi = 0, \quad J_{ab}(\L_b + T_b) \Phi =0,\quad (\L_a + T_a) \Phi =0.
\ee
To see the equivalence with the apparently stronger third condition in (\ref{equiv}) consider the identity $\xi_a^A P_{ab} (\L_b + T_b)\Phi = \xi^A_b J_b \Phi =0$, if we define $Y_a =\xi^a_b J_b$ one can prove from (\ref{LeftRight}) and (\ref{Adgm}) that $\{ Y_a \}$ generates the algebra $su(n+1)$. It is also known that the complement of the Lie subalgebra $s(u(n) \times u(1))$ spanned by $Y_A$ can generate the whole of $s(u(n) \times u(1))$ by commutation\footnote{ Such decomposition of the Lie algebra is called minimal \cite{Dmitri}, for the case at hand see \cite{Universal}.}, hence if $Y_A \Phi =0$ then $Y_{a}\Phi =0$ and it follows readily, contracting with $\xi^a_c$, that $J_c \Phi =0$. Also notice that the first and second conditions in (\ref{equiv}) are clearly equivalent.

\noindent Therefore we loose no generality in requiring the superpotential to satisfy 

\be \label{super}
(\L_a + T_a)\Phi =0.
\ee
We first observe that this system of differential equations is integrable, as can be seen by taking commutators $[J_a ,J_b]\Phi = if_{abc}J_c \Phi$, hence it has non-trivial solutions. 

However, if we were to demand that the superpotential be a function defined over $\CP^n$ globally, $\Phi(\xi)$, then there are only trivial solutions. This fact is a consequence of the $S(U(n)\times U(1) )$ representation content of $R$, indeed if we consider the equation (\ref{super}) with $\Phi (\xi)$ and project with $P_{ab}$ using $P_{ab}\L_b \Phi = \L_a \Phi$ we find

\be
(\delta_{ab} -P_{ab}) T_{b}\Phi (\xi) =0
\ee
and contraction with $T_a$ shows that the quadratic Casimir operator of $S(U(n)\times U(1))$ must vanish on $\Phi$

\be
T_{\ald} T_ {\ald} \Phi (\xi) = 0
\ee
now, since the Casimir operator is positive definite, this implies that when decomposing $\Phi$ into columns, seen as a matrix,  each column must be in the kernel of the forementioned Casimir operator, and this will produce non zero components only if the trivial representation is found in the reduction of $R$. The general reduction can be carried out straightforwardly following the branching rule presented in \cite{Huet}, and although we will not present the details here it is not difficult to prove that only the case $n=2$ contains the trivial representation ${\bf 1}_0$,

\be
R = {\bf 2}_3 \oplus  {\bf 3}_0 \oplus {\bf 1}_0 \oplus {\bf 2}_{-3}~,
\ee
where the subindices stand for the $U(1)$ charge. The analysis carried out already for $\CP^2$ in \cite{Huet} concludes that each column can only be a multiple of the harmonic spinor

\be
{\bf \Psi} = \sum_{k=1}^2  \xi_a  (\lambda_a)^{\alpha}_{\beta} \epsilon_{\alpha \mu \nu}e^{\mu \nu \beta}_k
\ee
and therefore the only possible non trivial solutions have the matrix form

\be 
\Phi (\xi) = (a_1 {\bf \Psi}, \cdots , a_{16} {\bf \Psi}), \quad \sum_{f= 1}^{16} |a_f|^2 = 1, \quad a_f \in \C
\ee
but such $\Phi$ is a singular matrix since its determinant obviously vanishes, hence it cannot be a superpotential, and we are led to point out the following 

\newtheorem*{R1}{Remark}

\begin{R1}
Global solutions of the superpotential with $\CP^n$ as domain are trivial.
\end{R1}
\noindent This is a manifestation of the known fact that a bundle admitting a global section is trivial.

If we extend our space of solutions to include functions on the bundle we see from proposition \ref{p8} that each column of $\Phi$ seen as a matrix belongs to $ \cap_a \ker( \L_a + T_a) = \mbox{im }\! \g$ and so the most general such solution is of the form $\Phi =\g \mathfrak{K}$, from the group property we must have $\mathfrak{K} \in Cliff(SU(n+1))$, concluding our proof of proposition \ref{p7} $\blacksquare \\$

It turns out that in our construction there is a universal superpotential from which the spin$_c$ connection may be derived in any spin$_c$ structure, and takes a particularly simple form, determined by the Clifford representation homomorphism $g \mapsto \g$. As noted this potential has a gauge freedom given by right group multiplication. Since $SU(n+1)$ is compact and connected the exponential map is surjective and we can give an explicit expression for our superpotential

\be
\Phi(g) = \prod_{w \in E} \mbox{exp}(-i \theta_w T_w) \mathfrak{K}
\ee
Where $\theta_w$ are the generalised Euler angles labeling $g$, which can be found explicitely in \cite{Todd}, $T_a$ are the generators of the Clifford representation and $E$ labels the subset of $su(n+1)$ having $n^2+2n$ elements required for the parametrisation. This provides us with a realisation for the superpotential on the bundle $SU(n+1) \to \CP^n$ corresponding to $U(1)$ charged spinor fields.

On the other hand we know that the choice of holomorphic or antiholomorphic coordinates is arbitrary to some extent, and this is patent in the $R$-symmetry of the superalgebra. Under a $U(1)$ rotation the supercharges will transform as 

\be
Ad(e^{i\theta N})[Q] = e^{i\theta} Q , \qquad Ad(e^{i \theta N})[Q^{\dagger}] = e^{-i\theta} Q^{\dagger}
\ee
such rotation will preserve the superalgebra relations and leave the superhamiltonian unchanged. Closed forms for the $R$-symmetry, or $U(1)$ holonomy, transformations are shown in appendix \ref{D} for $n=1,2,3,4$ along with the general procedure to calculate them for any $n \in \mathbb{N}$.

\section{Scope of the approach} \label{lastsec}

Even though we have dealt here only with the particular case of $\CP^n$, it is known that a wide class of symmetric riemannian spaces, seen as left coset spaces $G/H$, admit a spin structure and hence a straightforward Dirac operator \cite{Fried,Bal}. We show how to extend the present construction to spin manifolds $\mathcal{M} = G/H$ with $G$ and $H$ compact, connected, and $G$ semisimple, in fact we generalise proposition \ref{p7} to all such manifolds seen as base spaces over the principal bundle $G \longrightarrow^{\!\!\!\!\!\!\!\! \pi}~ \mathcal{M}$. The Lie algebra of $G$, $\mathfrak{a}$, decomposes naturally $\mathfrak{a} = \mathfrak{h} \oplus \mathfrak{m}$, where $\mathfrak{h}$ is the Lie algebra of $H$ and $\mathfrak{m}$ its Cartan orthogonal complement, in the general case we have

\[
[\mathfrak{m}, \mathfrak{m}]\subset \mathfrak{a},\quad  [\mathfrak{h}, \mathfrak{h}]  = \mathfrak{h}, \quad  [\mathfrak{h}, \mathfrak{m}]  \subset \mathfrak{m}
\]
The standard Dirac operator is $\gamma^A \K_A$, which under the local frame rotation of proposition \ref{p1} may be brought to the form

\be
\D = \gamma^{a} Ad_{Aa}(g) Ad_{Ab}(g) \left( \L_b + T_b \right) = -\g \circ \gamma^A \K_A \circ \g^{\dagger} 
\ee
and the covariant derivative is now $\nabla_a = Ad_{Aa}(g) Ad_{Ab}(g) \left( \L_b + T_b \right)$, since the adjoint representation is orthogonal\footnote{ This is the case for compact $G$, see e.g. \cite{ORaif} }we may also display the spin connection explicitely

\[
\D = \gamma^{a} (\L_a + Ad(g)_{ \Au b} \K_{\Au} + Ad_{Aa}(g)Ad_{Ab}(g)T_b)
\]
the action of $\K_{\Au}$ is lifted canonically to a spin rotation $\tau_{\Au} = (J_{\Au})_{ab} [\gamma^{a}, \gamma^b]$ by the spin structure through an adequate homomorphism $H \to Spin (|G/H|)$ , given by the generating matrices $J_{\Au}$, and hence the seemingly differential term above is actually algebraic on the spin bundle, contributing to the spin connection; this is the statement contained in theorem \ref{t1} for $\CP^n$. We can now give the explicit form of the spin connection on $\mathcal{M}$ with the general conditions given, in the left-action approach:

\be
\omega_{a} = Ad(g)_{\Au a} \tau_{\Au} + Ad(g)_{Ab} Ad(g)_{Aa} T_a 
\ee

\subsection{General left superpotential}

A superpotential is any function $\Phi : G \to Cliff $ that satisfies the defining relation $\nabla_a \Phi (g) = 0$, in particular notice $\nabla_a \g =0$. This suggests the change variables $\Phi (g) = \g \mathfrak{K} (g) $. We will prove that under very general conditions $\mathfrak{K} (g)$ is a constant element, rendering proposition \ref{p7} a particular case; it is not difficult to see that the condition is equivalent to $Ad_{A b } (g) \L_b \mathfrak{K}(g) =0$ and can be stated as

\be \label{KA}
\K_{A} \mathfrak{K}(g) = 0
\ee
this guarantees that $\mathfrak{K}|_{\mathcal{M}}$ is constant. We will prove that constancy on $\mathcal{M}$ implies constancy on the whole of $G$, i.e. that the property of covariant constancy can be lifted from $\mathcal{M}$ to $G$. To this end let $\check{g},\check{h}$ be the dual Coxeter numbers of $G$ and $H$ respectively \cite{Kac}, under the assumptions given for $G,H$ we have the relation

\be \label{KCu}
\K_{\Cu} = \frac{i}{2 (\check{h} - \check{g})} f_{AB \Cu} ([\K_{A}, \K_{B}] - i f_{ABC} \K_{C}) 
\ee
even when there is torsion; we remark that vanishing torsion is a weaker condition than minimality of the decomposition $\mathfrak{a} = \mathfrak{h} \oplus \mathfrak{m}$. From (\ref{KA}) and (\ref{KCu}) follows  $\K_{\Cu} \mathfrak{K} (g) = 0$, and hence $\K_a \mathfrak{K}(g) =0$, proving that $\mathfrak{K}$ is constant. It is now straightforward to show that no global superpotential exists with $\mathcal{M}$ as domain, for if this were the case then for any $h \in H$ we would have an equivariant function $\Phi (g) = \Phi (hg) = \g \mathfrak{K} = \mathfrak{H} \g \mathfrak{K} $, leading to conclude that $\mathfrak{H} = \mathfrak{e}$ which is a contradiction. Other ways to naturally extend this construction would be to gauge the Dirac operator with a Yang-Mills field, this was done e.g. in a local approach in \cite{Brian} with $H$ as the gauge group, or to consider generalised spin$_K$ structures \cite{Bal} such as that required for $SU(3)/SO(3)$, which needs $K=SU(2)$, where non-abelian background fields are required to even define spinors.

\section{Conclusions and Remarks}

We formulated the Dirac operator on $\CP^n = SU(n+1)/S(U(n) \times U(1))$ in terms of right-invariant Killing vector fields, instead of the usual left-invariant found in the current literature \cite{Fried,Bal} for a left coset; we found a universal Dirac operator for any spin$_c$ bundle. The problem was motivated by an ansatz for the Dirac operator proposed in \cite{Huet} which was found to describe the canonical spin$_c$ structure on $\CP^2$ and can now be seen as a particular case. We proved that our Dirac operator is unitarily equivalent to the left-invariant operator $\D_K$ through a point-dependent spinor frame rotation. The eigenvalue problem was completely solved and agreement was found with known results for the spectra. The construction was carried out in a global coordinate approach, without recourse to local methods. New expressions for the eigenspinors as well as the bundle curvature, spin connection and chirality were also given through a detailed analysis of the group representations involved.

Following the ideas in \cite{Wipf} we have shown how to realise the extended $\mathcal{N} =2$ supersymmetry algebra in our approach, which is relevant for the problem of a non-relativistic fermion confined to $\CP^n$, where the (super) hamiltonian is $H = \D^2$ and takes into account the $U(1)$ field introduced by the connection.

The general problem of finding a superpotential in our approach, $\Phi$, whose domain is the bundle was also solved in the case at hand $\CP^n$, and with a different technique in the more general setting of a coset space $G/H$ with spin structure where $G,H$ are compact, connected and $G$ semisimple, it was found that the general superpotential is essentially unique up to right translations and that a superpotential with the base space as domain is impossible.

\section*{Acknowledgements}

I. H. is indebted to A. Wipf for helpful discussions, to E. Wagner for enlightening observations and suggestions and to Conacyt for funding under the SNI program during the preparation of this article.
The support of COFAA (IPN, Mexico) is also acknowledged.

\begin{appendix}

\section{Normalisation of eigenspinors} \label{A}

In this appendix we find the normalisation prefactor $\nu_k (l,q,n)$ introduced in (\ref{eigenspin}); as a first step we compute the volume of $\CP^n$ in our conventions. Our starting point is the induced metric by the background space, after simplification using the Fierz identity and the constraint $|z|=1$:

\be
ds^2 = d\xi_a d\xi_a = \frac{2(n+1)}{n} (\dl^{\al}_{\bt} - z^{\al} \zb_\bt) d\zb_{\al}dz^{\bt}
\ee
The global coordinates $z$ have a $U(1)$ freedom, we use this freedom to choose a representative of each equivalence class in the coset in such manner that the last component is always 
real. This amounts to the parametrisation (valid on the open cover $\CP^n /O $ where $O$ is the subset where $z^{n+1}=0$):

\be
z^{\imath} = \frac{\zeta^{\imath}}{\sqrt{1 + |\zeta|^2}}, \qquad z^{n+1} = \frac{1}{\sqrt{1 + |\zeta|^2}}, \qquad |\zeta|^2 := \sum_{\imath =1}^n |\zeta^{\imath}|^2
\ee
These are the Fubini-Study coordinates for $\CP^n$, their range is the whole $\C^n$, after a short calculation we obtain the Fubini-Study metric:

\be \nn
ds^2 = \frac{2(n+1)}{n} \frac{1}{1+ |\zeta|^2} \Big( \dl^{\imath}_{\jmath} - \frac{\zeta^{\imath} \bar{\zeta}_{\jmath}}{1+ |\zeta|^2} \Big) d\bar{\zeta}_{\imath} d\zeta^{\jmath} 
= h_{\bar{\imath} \jmath} d\bar{\zeta}_{\imath} d\zeta^{\jmath} 
\ee
We have found that a convenient way to do this calculation is to split the differentials into

\be
dz^{\al} = \frac{(1- \dl^{\al}_{n+1})}{ \sqrt{1 + |\zeta|^2}  } d\zeta^{(\al)} - \frac{z^{\al}}{2} \frac{d(|\zeta|^2)}{1+|\zeta|^2 }
\ee
And observe that the terms containing a $z^{\al}$ or a $\zb_{\bt}$ will vanish by action of the projector $({\bf 1} - z \otimes \zb)$, the $(\al)$ indicates that this index is not summed 
and that it only makes sense for $\al = \jmath$. A straightforward calculation leads us to the determinant of the metric tensor $\det h = \Big(\frac{2(n+1)}{n} \Big)^{\!\! n} \frac{1}{(1+ |\zeta|^2)^{n+1}} $, therefore the volume form is \cite{Wells}

\be
\upsilon = \Big(\frac{2(n+1)}{n} \Big)^n \frac{1}{(1+ |\zeta|^2)^{n+1}} \Big(\frac{i}{2}\Big)^n \bigwedge_{\imath =1}^n dz^{\imath} \wedge d\zb_{\imath}
\ee
Observe that the set $O$ has zero Riemann measure, hence to find the volume we integrate over the open cover $\CP^n/O$

\be
\mbox{Vol}(\CP^n) = \int_{\CP^n /O} \upsilon = \Big(\frac{2(n+1)}{n} \Big)^n \int_{\C^n} \Big(\frac{i}{2}\Big)^n 
 \frac{1}{(1+ |\zeta|^2)^{n+1}}  \bigwedge_{\imath =1}^n dz^{\imath} \wedge d\zb_{\imath}
\ee
The integral in the r.h.s. is standard, it can be found e.g. in \cite{Kazuyuki} and has the value $\frac{\pi^n}{n!}$, therefore we find:

\be \label{volcpn}
\mbox{Vol}(\CP^n) = \frac{1}{n!} \Big( \frac{2(n+1)\pi}{n} \Big)^n
\ee
in the case $n=1$ it gives the correct value Vol$(S^2) = 4 \pi$. The eigenspinors will be normalised so that

\be \nn
\int_{\CP^n} \Psi^{(\sigma) \dagger} \Psi^{(\sigma')} \upsilon= \delta^{\sigma \sigma'} \Pi^{\bt_1 \cdots \bt_l ; \mu_1 \cdots \mu_{l+q}}_{\al_1 \cdots \al_{l+q}; \nu_1 \cdots \nu_{l}}
\ee
Where the indices hidden on the l.h.s. are explicit on the r.h.s., $\upsilon$ is the volume form of $\CP^n$ and $\Pi$ is the projector onto the appropriate representation: 

\be
\mbox{\raisebox{.65cm}{$\overbrace{\overline{\young(\ \Dots\ )}}^{l}$} } \hspace{-.215cm}
\overbrace{\young(\ \Dots\ ,\first,\vDots,\kth) }^{l+q-k}
\ee
The dimension of the representation is 

\be \label{dimrepn}
d = \frac{(l+n)! (l+q-k-1+n)! (2l+q-k+n)}{l!k!n! (l+q-k-1)! (n-k-1)!(l+n-k)(l+q)}
\ee
Representation theory leads to the normalisation condition being

\be
\frac{d}{4^k} = \nu_k(l,q,n)^2 \int_{\CP^n} \zb_{[ \al_0}\bar{u}^{\imath_1}_{\al_1} \cdots \bar{u}^{\imath_k}_{\al_k ]} z^{[\al_0} u^{\al_1}_{\jmath_1} \cdots u^{\al_k]}_{\jmath_k}
\langle \Omega | \Gamma^{\jmath_k} \cdots \Gamma^{\jmath_1} \Gamma^{\bar{\imath}_1} \cdots \Gamma^{\bar{\imath}_k}  \o \upsilon
\ee
and it is easy to show by direct computation that 

\be
\langle \Omega |  \Gamma^{\jmath_k} \cdots \Gamma^{\jmath_1} \Gamma^{\bar{\imath}_1} \cdots \Gamma^{\bar{\imath}_k}  \o = k! \delta^{[\jmath_1}_{\imath_1} \cdots \delta^{\jmath_k]}_{\imath_k}
\ee
Therefore we have

\bear
\frac{d}{4^k} &=& \nu_k(l,q,n)^2 k! \int_{\CP^n} \zb_{[ \al_0}\bar{u}^{\imath_1}_{\al_1} \cdots \bar{u}^{\imath_k}_{\al_k ]} z^{[\al_0} u^{\al_1}_{\imath_1} \cdots u^{\al_k]}_{\imath_k}  \upsilon
\\ \nn
&=&\nu_k(l,q,n)^2 k! \int_{\CP^n} \zb_{ \al_0}\bar{u}^{\imath_1}_{\al_1} \cdots \bar{u}^{\imath_k}_{\al_k } z^{[\al_0} u^{\al_1}_{\imath_1} \cdots u^{\al_k]}_{\imath_k}  \upsilon \\ \nn
&=& \nu_k(l,q,n)^2 k! \frac{1}{(k+1)} \int_{\CP^n} \zb_{ \al_0}\bar{u}^{\imath_1}_{\al_1} \cdots \bar{u}^{\imath_k}_{\al_k } z^{\al_0} u^{[\al_1}_{\imath_1} \cdots u^{\al_k]}_{\imath_k} 
\upsilon  \\ \nn
&=& \frac{\nu_k(l,q,n)^2 k!}{k+1} \int_{\CP^n} \bar{u}^{\imath_1}_{\al_1} \cdots \bar{u}^{\imath_k}_{\al_k } u^{[\al_1}_{\imath_1} \cdots u^{\al_k]}_{\imath_k}  \upsilon 
\ear
The integrand can be evaluated using combinatorics, it is the trace of an antisymmetrizer with $k$ indices where each index runs from 1 to $n$, i.e. $\binom{n}{k}$ (The contractions of 
$u$'s give a rank $n$ projector instead of a Kronecker delta), hence we have:

\be
\frac{d}{4^k} = \frac{\nu_k(l,q,n)^2 k!}{k+1} \binom{n}{k} \mbox{Vol}(\CP^n)
\ee
finally using (\ref{volcpn}) one finds

\be
\nu_k(l,q,n) = 2^{-k} \sqrt{d(k+1)(n-k)!} \Big( \frac{n}{2(n+1)\pi} \Big)^{\frac{n}{2}}
\ee

\section{Spectrum} \label{B}

In this appendix we present the spectrum calculation of the proposed left action universal Dirac operator. This spectrum is not new, as it was found quite generally in \cite{Universal}, but we show how to compute it within the left action approach. First, we rewrite the spin laplacian as

\be
D^2 = J^2 - J_I^2 - (\xi \cdot J)^2
\ee
where we have defined $J_I : = \xi_a^I J_a$. Theorem \ref{t1} allows us to find the action of the $U(1)$ part above, by the use of (\ref{phiom}) and the following replacement rule, valid within the expression for the eigenspinors (\ref{eigenspin2}):

\be
[\phi, \bar{z}_{\gamma} (\lambda_a)^{\gamma}_{\alpha}\gamma^a ] \to  -\frac{2}{n} \bar{z}_{\gamma}(\lambda_a)^{\gamma}_{\alpha} \gamma^a 
\ee
which leads readily to 

\be
\phi \Psi^{(\sigma)} = \Big(\frac{n-2k}{n} \Big) \Psi^{(\sigma)}
\ee
the remaining piece needed to compute the eigenvalue of the spin laplacian is the action of the Casimir operator $\tau_I^2$; to this end it is more convenient to introduce the tensors $\xi^{\bar{\imath}}_a = \bar{z}_\alpha\frac{(\lambda_a)^{\alpha}_{\beta}}{2}u^{\beta}_{\imath} $, defined through $\g \Gamma^{\bar{\imath}} \g^{\dagger} = \xi^{\bar{\imath}}_{a} \gamma^a$,  and rewrite (\ref{eigenspin}) in the form:

\be
\Psi^{(\sigma)} = 2^k \nu_k(l,q,n) \bar{z}_{\alpha_{k+1}} \cdots \bar{z}_{[\alpha_{l+q}}\bar{u}_{\alpha_1}^{\imath_1} \cdots \bar{u}_{\alpha_k]}^{\imath_k}z^{\beta_1} \cdots z^{\beta_l} \xi_{a_1}^{[\bar{\imath}_1} \cdots \xi_{a_k}^{\bar{\imath}_k ]} \gamma^{a_1} \cdots \gamma^{a_k}\om
\ee
Notice $[\tau_I , \xi^{\bar{\imath}}_a \gamma^a ] = \frac{(\lambda_I)^{\jmath}_{\imath}}{2} \xi^{\bar{\jmath}}_a \gamma^a$, so that $\Psi^{(\sigma)}$ carries a totally antisymmetric $SU(n)$ irreducible representation. The problem reduces to evaluating Casimir operators for the relevant representations, this calculation could also be done in a somewhat simpler way using proposition \ref{p3}, following our conventions \cite{Perelomov} we obtain 

\be
C_2^{su(n+1)}[B_1, \cdots, B_n] = \half \left[ \sum_{p=1}^{n}B_p^2 - \frac{ \left(\sum_{p=1}^{n} B_p\right)^2}{n+1} - 2 \sum_{p=1}^n pB_p + (n+2) \sum_{p=1}^{n} B_p \right]
\ee
In the expression above the Casimir is defined as the sum of squares of our generators, and the notation $[B_1, \cdots, B_n]$ is the box number labeling for Young tableaux, where we chose $B_1 \geq B_2 \geq \cdots \geq B_n$. The eigenvalue of the eigenspinor $\Psi^{(\sigma)}$ associated to $J^2$ is that of $[2l+q-k,\underbrace{l+1,\cdots, l+1}_{k}, \underbrace{l, \cdots, l}_{n-k-1}]$ and for $\tau_I^2$ is that of $[\underbrace{1, 1, \cdots, 1}_{k}, \underbrace{0, \cdots, 0}_{n-k-1}]$, hence we find

\vspace{0.3cm}

\be
D^2 \Psi^{(\sigma)} = \left(  l^2 + l(q+n-k) + \frac{qn-k(n+1)}{2} \right) \Psi^{(\sigma)}
\ee
Finally, for the curvature contribution observe that one can rewrite, on account of (\ref{curvident})

\be
\frac{\gamma \cdot F}{2} = \frac{\sqrt{n(n+1)}}{\sqrt{2}} \phi ~\xi \cdot J - 2 \tau_I J_I,
\ee
use then theorem \ref{t1} and proposition \ref{p2} again to find:

\be
\D^2 \Psi^{(\sigma)} = (l+q)(l+n-k)\Psi^{(\sigma)} = \Lambda_{\sigma} \Psi^{(\sigma)}
\ee
The degeneracy of each eigenvalue is the dimension of the $SU(n+1)$ irreducible representation that labels $\Psi^{(\sigma)}$, i.e. $\mbox{deg}(\Lambda_{\sigma}) = d$, given in (\ref{dimrepn}). As we mentioned, the non zero portion of the spectrum is given by $\{ \pm \sqrt{\Lambda_{\sigma}} \}$, while the zero modes exist only when either $q \leq 0$ or $q \geq n+1$ and have degeneracy given by the dimension of the corresponding irreducible representations, namely deg$(\chi) = \binom{|q|+n}{n}$ and deg$(\psi) = \binom{q-1}{n}$. These formulae conform to known results \cite{Universal,Cahen,Seifarth,Baer}.

The index of the Dirac operator on $\CP^n$ should be \cite{Nash,Smilga}

\be
\mbox{ind}(\D) = \frac{(1-q) \cdots (n-q)}{n!}.
\ee
For $q \leq 0$ the index coincides with deg$(\chi)$ while for $q \geq n+1$ it is the same as $(-1)^n$deg$(\psi)$ and vanishes when $0< q < n+1$. Hence we see that $\D$ has the correct Atiyah-Singer index.

\section{Reduction of the Clifford representation} \label{C}

Here we show that the representation $Cliff$ is a direct sum of identical copies of the representation $(1,1, \cdots, 1)$. In order to do so we will show that their highest weight vectors 
are identical. Let $m = 1, 2, \cdots, n$ and label by $T_{ \{m \} }$ the elements of the Cartan subalgebra of $su(n+1)$ in the Gell-Mann basis, that is $\{m\} =m^2 +2m$. By the very definition we have: $T_{ \{m \} }\o = \mu_{m}^{Cliff} \o$ where $\mu^{Cliff}$ is the weight vector corresponding to $\o$. In this last section we do not use the index conventions set in the rest of the article, the conventions used here are clear from the context or explained herein. Einstein's summation convention will not be used in this section. Our main goal is to compute the weight vector of $\o$, to achieve this we consider first

\be
T_{ \{m \} } \o = \frac{1}{4i} \sum_{a,b =1}^{ \{n\} } f_{\{ m\} ab} \gm^{ab} \o
\ee
It is obvious that $f_{ \{m_1\} \{m_2\} a } = 0$ (by definition of Cartan subalgebra). The vacuum is defined by the property $ \Gamma^{\J} \o =0 $. We reorder the labels $\J$ in such a way that the values $\J = 1, \cdots, \frac{n(n+1)}{2} $ correspond to the holomorphic complement of the Cartan subalgebra, for these values we shall use the indices $i,j$ instead. We will also use the shorthand $\frac{n(n+1)}{2}=M$. Corresponding to a holomorphic gamma matrix of the form $\Gamma^{i} = \half (\gm^{s} - i \gm^{s+1})$ we associate the matrices $\lm_{i} =  \half(\lm_{s} + i \lm_{s+1})$, $\lm_{\bar{i}} =  (\lm_{i})^{\dagger} $ and define the complexified structure constants accordingly:

\be
T_{\{m\}} \o = \frac{1}{4i} \sum_{i,j=1}^{M} (f_{ \{m\} i\bar{j} } \Gamma^{i} \Gamma^{\bar{j}} + f_{ \{m\} \bar{i} \bar{j} } \Gamma^{\bar{i}} \Gamma^{\bar{j}} ) \o
\ee
a short calculation yields

\be
\half \sum_{i=1}^{M} [ \lm_{i}, \lm_{\bar{i}}] = \frac{1}{4i} \sum_{i=1}^{M} \sum_{a=1}^{ \{n \} } \lm_a f_{a i \bar{i}}
\ee
It is not difficult to see that $f_{\{ m\} \bar{i} \bar{j} } = 0$, the reason is that the commutator of two strictly upper triangular elementary matrices has zero diagonal. Combining 
these results it follows that 

\be \label{Tm}
T_{\{m\}}\o = \sum_{i=1}^M \frac{f_{ \{m \} i \bar{i} } }{4i} \o =  \half \sum_{i=1}^M \tr \Big( \frac{\lm_{ \{ m\} }}{2} [\lm_{i}, \lm_{\bar{i}} ] \Big) \o
\ee
Now we look at a general $\lm_{i}$, it has matrix elements given by some choice of the pair $(l,k)$

\be
(\lm_{i})^{\al}_{\bt} = \dl^{\al}_k \dl^{l}_{\bt}, \qquad k < l, \quad 1<l, \quad k< n+1, \quad \al, \bt = 1, \cdots, n+1
\ee
Working out the commutator in (\ref{Tm}) one gets a diagonal matrix with elements

\be
\sum_{i=1}^M ([\lm_{i}, \lm_{\bar{i}}])^{\al}_{\al} = \sum_{l=2}^{n+1} \sum_{k<l} (\dl^{\al}_{k} - \dl^{\al}_{l})
\ee
Also useful is the expression

\be
 \frac{(\lm_{\{m\}})^{\al}_{\al} }{2} = \frac{1}{\sqrt{2m(m+1)}} \bigg( \sum_{j=1}^{m} \dl^{j}_{\al} - m \dl^{m+1}_{\al} \bigg)
\ee
In calculating the trace only the following sums are necessary

\be
\sum_{l=2}^{n+1} \sum_{k<l} \dl^{j}_{k}  = n+1-j, \quad \sum_{l=2}^{n+1} \sum_{k<l} \dl^{j}_{l} = j-1
\ee
Using these results it follows readily that

\be
\tr \Big( \frac{\lm_{ \{ m\} }}{2} [\lm_{i}, \lm_{\bar{i}} ] \Big) = \frac{1}{\sqrt{2m(m+1)}}\bigg( \sum_{j=1}^{m} ( n+2 - 2j) - m (n-2m) \bigg)
\ee
and hence

\be
T_{\{m\} }\o =  \frac{\sqrt{m(m+1)}}{\sqrt{8}} \o = \mu^{Cliff}_m \o
\ee
This can be compared, in the same conventions, with the highest weight of the representation $R:= (1, 1, \cdots, 1)$ of $su(n+1)$ (all Dynkin indices are one), this weight is well known \cite{Perelomov}

\be
\mu^{R}_m = \frac{1}{\sqrt{2m(m+1)}}\sum_{k=1}^{n} \sum_{l=1}^{k} \bigg( \sum_{j=1}^m \dl^{j}_{l} - m \dl^{m+1}_{l} \bigg)
\ee
The relevant sum needed is 

\be
\sum_{k=1}^n \sum_{l=1}^k \dl^{j}_{l} = n+1-j
\ee
Finally, we obtain directly

\be
\mu^R_{m} = \frac{1}{\sqrt{2m(m+1)}} \Big( \sum_{j=1}^m (n+1-j) - m(n-m) \Big) = \frac{\sqrt{m(m+1)}}{\sqrt{8}}  =  \mu^{Cliff}_m
\ee
This shows that $\o$ is the highest weight state of $R$ in the conventions of Georgi \cite{Perelomov}, and this is only possible if the Clifford representation decomposes into a sum of identical copies:

\be
Cliff = \bigoplus_{r=1}^{  2^{ \lceil \frac{n}{2} \rceil }} R
\ee
The multiplicity is obtained from $dim(R) = 2^{n(n+1)/2}$ and the dimension of the background space which is, taking into account the possible addition of an extra gamma 
matrix, $2(n + \lceil n^2/2 \rceil) $.

\section{$U(1)$ holonomy transformations} \label{D}

We consider the non-trivial part of a general $R$-symmetry transformation $S_n(\theta) = e^{-i\frac{\phi n}{2} \theta}$. In this last section we abandon our index convention for greek indices while the symbol $\lambda$ will denote an eigenvalue. It will be convenient to use $\varphi = \frac{n}{2}\phi$. It was shown in \cite{Huet} that the minimum polynomial of $\varphi$ can be obtained by considering the invariants\footnote{Notice that our normalisation for $\phi$ differs from that in \cite{Huet}, $\varphi$ was called $\phi$ there since $n =2$.}

\be
\mathfrak{I}_k  = \frac{1}{(4i)^k}J_{a_1 b_1} \cdots J_{a_k b_k} \gamma^{a_1 b_1 \cdots a_k b_k}~,
\ee
being $\mathfrak{I}_1 = \varphi$, $\mathfrak{I}_{0}=1$ and $\mathfrak{I}_k = 0$ if $k \geq n+1$. These invariants are not independent but satisfy the recursive relations

\be
\mathfrak{I}_{k+1} = \mathfrak{I}_k \mathfrak{I}_1 + a_k \mathfrak{I}_{k-1}~,
\ee
with $a_k = - k(n+1-k)/4$. Since the system terminates at $k=n+1$ and all invariants are linearly independent we obtain the minimum polynomial of $\varphi$, called $p_n(\varphi)=\sum_m c_m^{(n)} \varphi^m$. Inspection shows that $p_n$ has degree $n+1$, parity $(-1)^{n+1}$ and $c_{n+1}^{(n)}=1$. To find a closed expression for the coeffcients from the recursion relation is an intricate combinatorial problem, which we shall solve in a different manner using repsentation theory. In the combinatorial approach $c^{(n)}_m$ could be calculated as a sum over paths on a lattice, leading for $q \geq 1$ to

\be \label{cn}
c_{n+1 -2q}^{(n)} = \sum_{(\al_j) \in C_q} a_{\al_1} \cdots a_{\al_q}
\ee
where the sum runs over all vectors $(\al_j) = (\al_1, \cdots , \al_q)$ in the set (where $  1 \leq j \leq q$)

\be
C_q = \left\{ \left( n-\sum_{k=1}^{j}i_k - 2(j-1) \right) ~~:~~ \sum_{k=1}^{q+1-h}i_k + 2q = n + h, ~i_k \geq 0, ~ h =0,1 \right\},
\ee
this set has $|C_q| = \binom{n+1-q}{q}$. Using this expression for $p_n$ repeatedly we could cast $S_n(\theta)$ into polynomial form by brute force, however such approach is rather cumbersome and not necessary, as we can find all the roots of the minimal polynomial from representation theory, which simplifies notably the task of reaching such polynomial form. To this end consider the action of $\varphi$ on an eigenvector in the $Cliff$ representation, such vector is a direct sum of tensors that correspond to standard fillings of the tableau for $R$ sharing the same eigenvalue. In calculating the spectrum we need only consider fillings that give different eigenvalues. When giving a filling, the number $n+1$, if present in the filling, could only occupy the lowest box in a column, this means that the number of all boxes filled with $n+1$, which we call $l$, can only be $0 \leq l \leq n$. From here we see that the $\varphi$ eigenvalue receives the same contribution from all other boxes and the only negative contribution from these $l$ boxes, these charges are in the relation $1:-n$, (seen from the fundamental representation), thus the eigenvalue $\lambda_l$ depends only on $l$, taking into account normalisations:

\be
\lambda_l = \frac{1}{n+1}\left( \frac{n(n+1)}{2} - l  - l n \right) = \frac{n}{2} - l, \qquad 0 \leq l \leq n.
\ee
Let $P_l$ denote the projector onto the eigenspace of $\lambda_l$, then the spectral theorem gives:

\be \label{specthm}
S_n(\theta) = e^{-i\theta \varphi} = \sum_{l=0}^n e^{-i\theta \lambda_l} P_l,\qquad  P_l = \prod_{ r \neq l} \left(\frac{\varphi - \lambda_r}{\lambda_l - \lambda_r} \right), \quad p_n(\varphi) = \prod_{l=0}^n (\varphi -\lambda_l)~.
\ee
We list the $U(1)$ holonomy transformations on spinors for the first four complex projective planes:

\bear \nn
S_1(\theta) &=& \cos(\theta/2) -2i \varphi \sin(\theta/2)~, \\ \nn 
S_2(\theta) &=&  1 -\varphi^2 + \varphi^2 \cos \theta - i \varphi \sin \theta~,   \\ \nn
S_3(\theta) &=& \bigg(\frac{9}{4}-\varphi^2  \bigg) \left( \half \cos(\theta/2) - i \varphi \sin (\theta/2) \right) + \bigg(\varphi^2 -\frac{1}{4} \bigg)\bigg(\half \cos(3\theta/2)- \frac{i}{3} \varphi \sin(3\theta/2) \bigg)~.      \\ \nn
S_4(\theta) &=&\frac{(\varphi^2 -1)(\varphi^2 -4)}{4} + \frac{\varphi (4-\varphi^2)}{3} (\varphi \cos \theta - i \sin \theta) + \frac{\varphi (\varphi^2 -1)}{12}(\varphi \cos(2\theta) -2i \sin(2\theta)).
\ear
For completeness we give a compact expression for the coefficients $c_m^{(n)}$ in (\ref{cn}) in terms of the unsigned Stirling numbers of the first kind $n\brack m$ from (\ref{specthm}):

\be
c_m^{(n)} = \sum_{k=m}^{n+1} {n+1 \brack m} \binom{k}{m} (-1)^{k-m} \left(\frac{n}{2} \right)^{k-m}
\ee

\end{appendix}
%
%
\bibliographystyle{abbrv}

\begin{thebibliography}{10}
\bibitem{Grundland} Grundland, A. M., and S. Post, ``{\it Soliton surfaces associated with $\CP^{N-1}$ sigma mo\-dels.}'' Journal of Physics: Conference Series. Vol. \textbf{380.} No. 1. IOP Publishing, 2012, [math-ph/1112.2420]
\bibitem{AAWipf} Aguado, Miguel, M. Asorey, and A. Wipf, ``{\it Nahm Transform and Moduli Spaces of $\CP^n$-Models on the Torus.}'' Ann. Phys. \textbf{298.1} (2002): 2-23, [hep-th/0107258]
\bibitem{Rajamaran} Rajaraman, R. ``{\it ${\bf CP}_N$ Solitons in Quantum Hall systems}'', [cond-mat/0112491]
\bibitem{KarNair} Karabali, Dimitra, and V. Parameswaran Nair, ``{\it Quantum Hall effect in higher dimensions.}'' Nucl. Phys. \textbf{B 641.3} (2002): 533-546, [hep-th/0203264]
\bibitem{Seifarth} S. Seifarth, U. Semmelmann, ``{\it The spectrum of the Dirac operator on complex projective spaces}'', SFB 288 preprint no. \textbf{85}, Berlin (1993), [math/9801091]
\bibitem{Baer} Ammann, Bernd, and Christian B\"ar. ``{\it The Dirac operator on nilmanifolds and collapsing circle bundles.}'' Annals of Global Analysis and Geometry \textbf{16.3} (1998): 221-253, [math/98101091]
\bibitem{Wipf} Kirchberg, A., J. D. L\"ange, and A. Wipf. ``{\it Extended supersymmetries and the Dirac operator.}'' Annals of Physics \textbf{315.2} (2005): 467-487, [hep-th/0401134]
\bibitem{Universal} Dolan, Brian P., Huet I., Murray S., O'Connor D. ``{\it A universal Dirac operator and noncommutative spin bundles over fuzzy complex projective spaces.}'' Journal of High Energy Physics \textbf{2008.03} (2008): \textbf{029}, [hep-th/0711.1347]
\bibitem{Huet} Huet, I. ``{\it A projective Dirac operator on $\CP^2$ within fuzzy geometry.}'' Journal of High Energy Physics \textbf{2011.2} (2011): 1-28, [hep-th/1011.0647]
\bibitem{Smilga} Ivanov, E. A., and A. V. Smilga. ``{\it Dirac operator on complex manifolds and supersymmetric quantum mechanics.}'' International Journal of Modern Physics A \textbf{27.25} (2012), [hep-th/1012.2069]
\bibitem{Habib} Habib, Georges, and Roger Nakad. ``{\it The twisted Dirac operator on K\"ahler submanifolds of the complex projective space.}'' Journal of Geometry and Physics \textbf{77} (2014): 43-47, [math/1207.2642]
\bibitem{Bal} A. P. Balachandran, G. Immirizi, J. Lee, P. Pre\v snajder, ``{\it Dirac operators on coset spaces.}'', J. Math. Phys. \textbf{ 44} (2003) 4713-4735, [hep-th/0210297]
\bibitem{Fried} Friedrich, Thomas ``{\it Dirac operators in Riemannian geometry.}'' Vol. \textbf{25}. p. 82-88, Providence: American Mathematical Society, 2000.
\bibitem{Nakad} Nakad, Roger, and Mihaela Pilca. ``{\it Eigenvalue estimates of the spin$^c$ Dirac operator and harmonic forms on k\"ahler-Einstein manifolds.} (2015), [math/1502.05252]
\bibitem{Brian} Dolan, Brian P. ``{\it The spectrum of the Dirac operator on coset spaces with homogeneous gauge fields.}'', Journal of High Energy Physics \textbf{2003.05} (2003): 018, [hep-th/0304037]
\bibitem{BDolan} Dolan, Brian P. ``{\it Non-commutative complex projective spaces and the standard model.}'' Mod. Phys. Lett. A \textbf{18.33} no. 35 (2003): 2319-2327, [hep-th/0307124]
\bibitem{susycpn} Bellucci, S., et al. ``{\it Symmetries of $\mathcal{N}= 4$ supersymmetric $\mathbb{CP}^n$ mechanics.}'' Journal of Physics A: Mathematical and Theoretical \textbf{46.27} (2013): 275305, [hep-th/1206.0175], Candu, Constantin, et al. ``{\it The sigma model on complex projective superspaces.}'' Journal of High Energy Physics \textbf{2010.}2 (2010): 1-48., Murray, Se\'an, and Christian S\"amann, ``{\it Quantization of flag manifolds and their supersymmetric extensions.}'' Advances in Theoretical and Mathematical Physics 12.3 (2008): 641-710 [hep-th/0611328]
\bibitem{CPNF} A. P. Balachandran, B. P. Dolan, J. H. Lee, X. Martin and D. O'Connor, ``{\it Fuzzy complex projective spaces and their star products}'', {\it J. Geom. Phys.} \textbf{ 43} (2002) 184, [hep-th/0107099]
\bibitem{Eguchi} Eguchi, Tohru, Peter B. Gilkey, and Andrew J. Hanson. ``{\it Gravitation, gauge theories and differential geometry.}'' Physics reports \textbf{66.6} (1980): 213-393.
\bibitem{Nomizu} Kobayashi, Shoshichi, and Katsumi Nomizu. ``{\it Foundations of differential geometry}''. Vol.\textbf{ 2}. XI, p. 273, Interscience Publishers, 1969.
\bibitem{ORaif} O'Raifeartaigh, Lochlainn, ``{\it Group structure of gauge theories.}'' (Cambridge University Press, 1988) p. 26.
\bibitem{Baer2} C. B\"ar, ``{\it Dependance on the spin structure of the Dirac spectrum}'', Seminaires et Congres,\textbf{ 4}, Global Analysis and Harmonic Analysis, J.P. Bourguignon, T. Branson, O. Hijazi (Eds.), 17-33 (2000), [math/0007131].
\bibitem{Cahen} M.Cahen, A. Franc and S. Gutt, ``{\it Spectrum of the Dirac operator on complex projective space $\mathbb{P}_{2q-1}(\mathbb{C})$}'', Lett. Math. Phys. {\bf 18} (1989) 165, M.Cahen, A. Franc and S. Gutt, ``{\it Erratum to `Spectrum of the Dirac operator on complex projective space $\mathbb{P}_{2q-1}(\mathbb{C})$'}'' Lett. Math. Phys. {\bf 32} (1994) 365.
\bibitem{GSW} M. B. Green, J. H. Schwarz, E. Witten ``{\it Superstring theory}'' Vol. 2, (Cambridge University press 1999) p. 445.
\bibitem{Witten} Witten, Edward, ``{\it Dynamical breaking of supersymmetry.}'' Nuclear Physics B \textbf{188.3} (1981): 513-554.
\bibitem{Lawson} H. B. Lawson Jr. and Marie-Louise Michelsohn,``{\it Spin geometry}'', (Princeton mathematical series {\bf 38}, 1989) p. 391-392. 
\bibitem{Dmitri} Cort\'es, Vicente, ed. ``{\it Handbook of pseudo-Riemannian geometry and supersymmetry.}'' Vol.\textbf{ 16} (European Mathematical Society, 2010) p.706.
\bibitem{Todd} Tilma, Todd, and E. C. G. Sudarshan. ``{\it Generalized Euler angle parametrization for $SU(N)$.}'' Journal of Physics A: Mathematical and General \textbf{35.48} (2002): 10467, [math-ph/0205016]
\bibitem{Kac} V. G. Kac, ``{\it Infinite dimensional Lie algebras}'', Cambridge University Press (1990), p. 80.
\bibitem{Wells} Wells, Raymond O'Neil. ``{\it Differential analysis on complex manifolds.}'' Vol.\textbf{ 65} (Springer, 2008) p. 190.
\bibitem{Kazuyuki} Fujii, Kazuyuki. ``{\it Introduction to Grassmann manifolds and quantum computation.}'' Journal of Applied Mathematics \textbf{2.8} (2002): 371-405, [quant-ph/0103011]
\bibitem{Perelomov} A. Perelomov and V. Popov ``{\it Casimir operators for $U(N)$ and $SU(N)$}'', Soviet Journal of Nuclear Physics, Vol. \textbf{3}, Number 5, 676-680 (1966), Georgi, H. ``{\it Lie algebras in particle physics}'', (Perseus 1999) p.188-191.
\bibitem{Nash} Dolan, Brian P., and Charles Nash. ``{\it The standard model fermion spectrum from complex projective spaces.}'' Journal of High Energy Physics \textbf{2002.10} (2002): 041.
\end{thebibliography}

\end{document}